\documentclass[aps, pre,twocolumn,showpacs,preprintnumbers, amsmath,amssymb]{revtex4}

\usepackage{graphicx}
\usepackage{ifthen}

\begin{document} \title{Instability of flat disks with respect to the formation of  twisted ribbons in smectic-\textit{A}$^*$ monolayers}

\author{Hao Tu and Robert A. Pelcovits} \affiliation{Department of Physics, Brown University, Providence RI, 02912, U.S.A} \date{\today}

\begin{abstract} Smectic-\textit{A}$^*$ monolayers self-assembled from aqueous solutions of chiral \textit{fd} viruses and a polymer depletant can assume a variety of shapes such as flat disks and twisted ribbons. A first order phase transition from a flat disk to a ribbon occurs upon lowering the concentration of polymer depletant or the temperature. A theoretical model based on the de Gennes model for the smectic \textit{A} phase, the Helfrich model of membrane elasticity and a simple edge energy has been previously used to calculate the disk-ribbon phase diagram. In this paper we apply this model to the nucleation process of ribbons. First, we study the ``rippled disks'' that have been observed as precursors of ribbons. Using a model shape proposed by Meyer which includes rippling in both the in-plane and out of plane directions, we study the energetics of the disks as functions of the edge energy modulus (a measure of the polymer concentration) and the mean curvature modulus $k$. We find that as the edge energy modulus is reduced the radial size of the ripples grows rapidly in agreement with experimental observations.  For small enough $k$ we find that the out of plane size of the ripples grows  but its value saturates at a fraction of the twist penetration depth, too small to be experimentally observable. For large $k$ the membrane remains flat though rippled in the radial direction. Such membranes do not have negative Gaussian curvature and thus will not likely spawn twisted ribbons. We also study the creation of twisted ribbons produced by stretching the edge of a flat membrane in a localized region. In experiments using a pair of optical traps it has been observed that once the membrane has been sufficiently stretched a ribbon forms on the stretched edge. We study this process theoretically using a free energy consisting of the Helfrich and edge energies alone. We add a small ribbon-like perturbation to the protrusion producd by stretching and determine whether it is energetically favorable as a function of the size of the protrusion. In qualitative agreement with experiment we find a nonzero value for the critical size of the protrusion needed to make a ribbon energetically favorable, though the value we find is an order of magnitude lower than the experimental value possibly due to our neglect of the director field. As in the case of the rippled disks 
we find that the mean curvature energy acts as a barrier between the disk and twisted ribbon structures.
\end{abstract}


\maketitle

\section{INTRODUCTION}
\label{intro}

\begin{figure}
\centering
\includegraphics[width=0.5in]{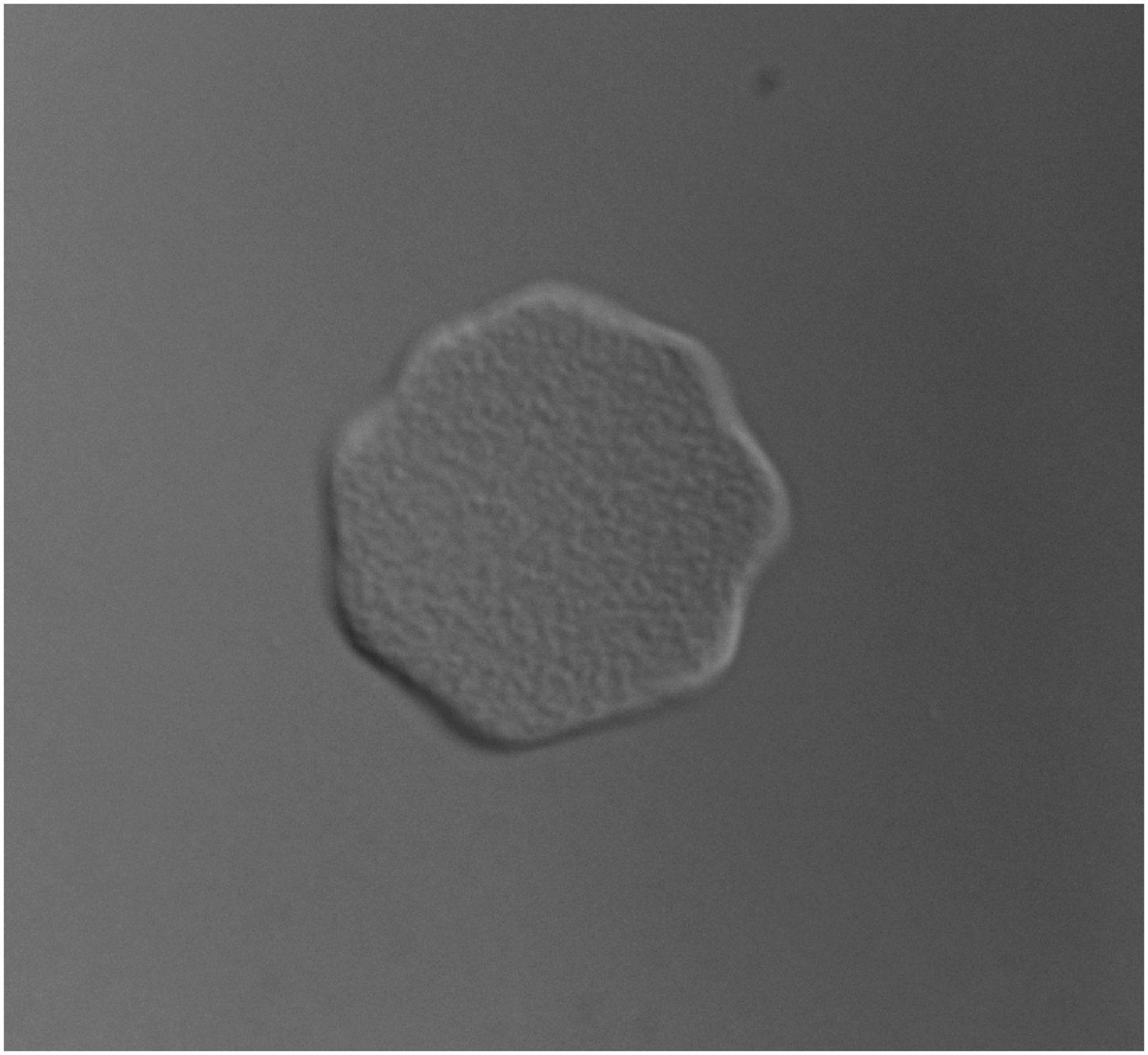}
\includegraphics[width=0.5in]{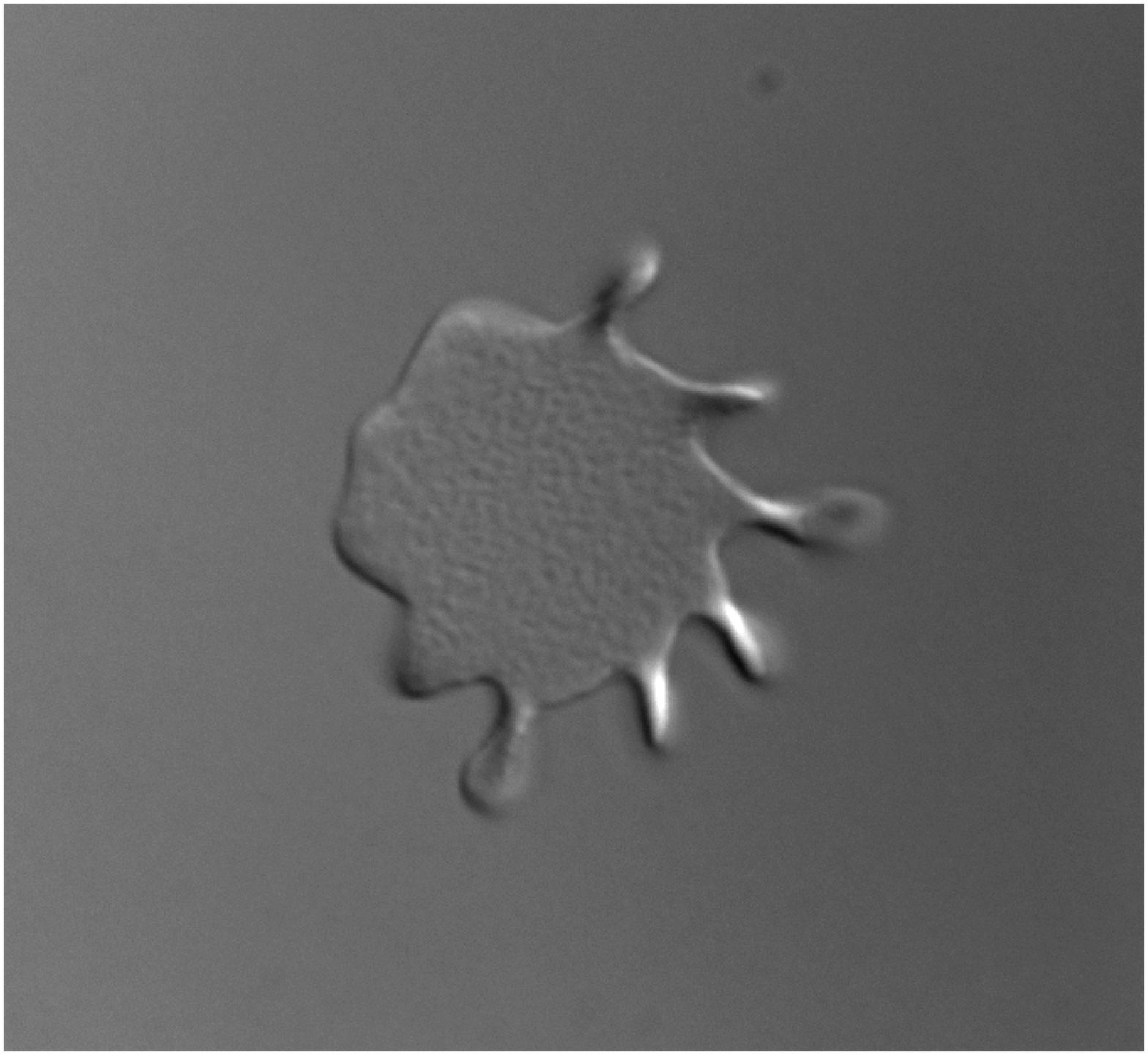}
\includegraphics[width=0.5in]{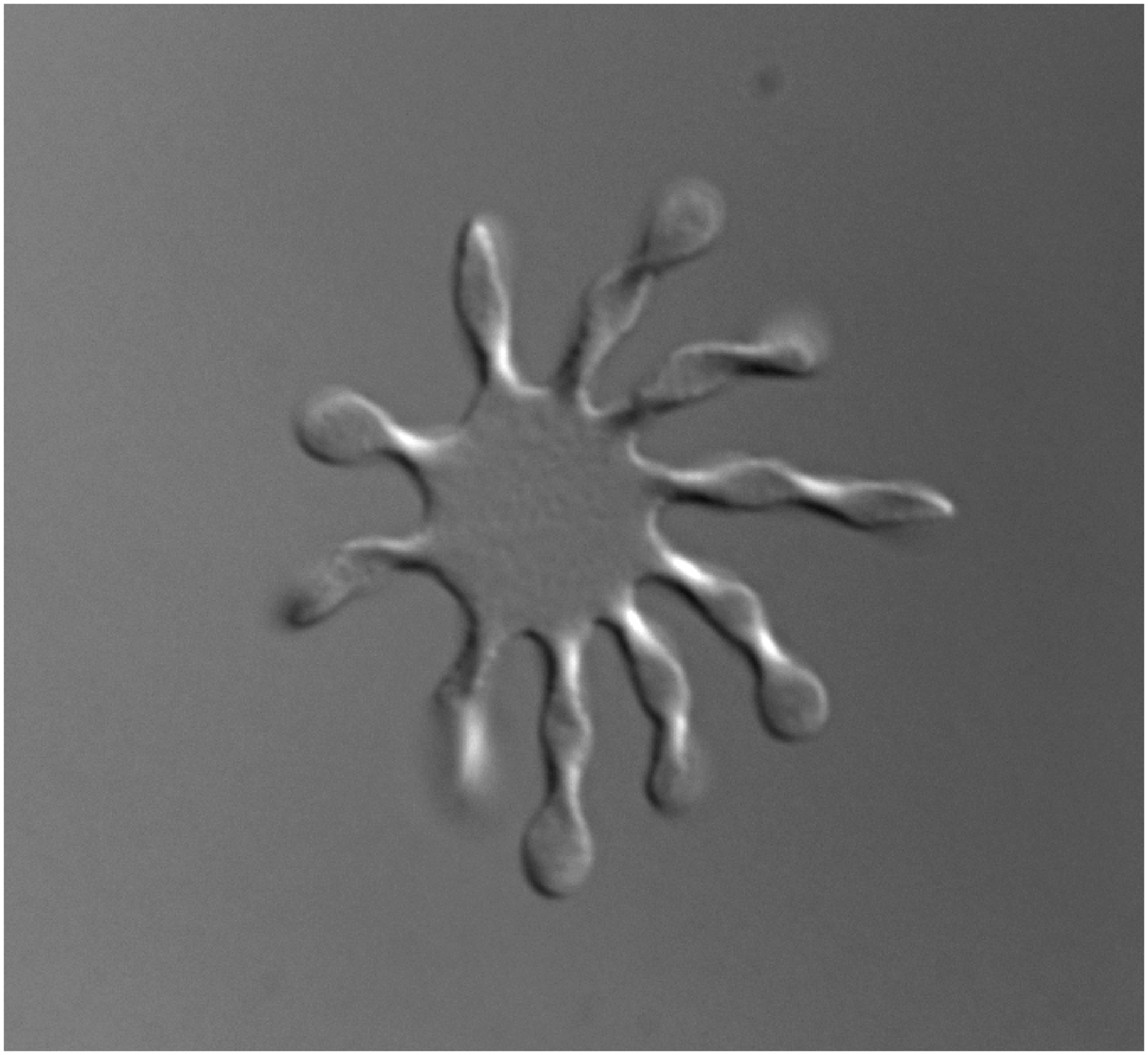}
\includegraphics[width=0.5in]{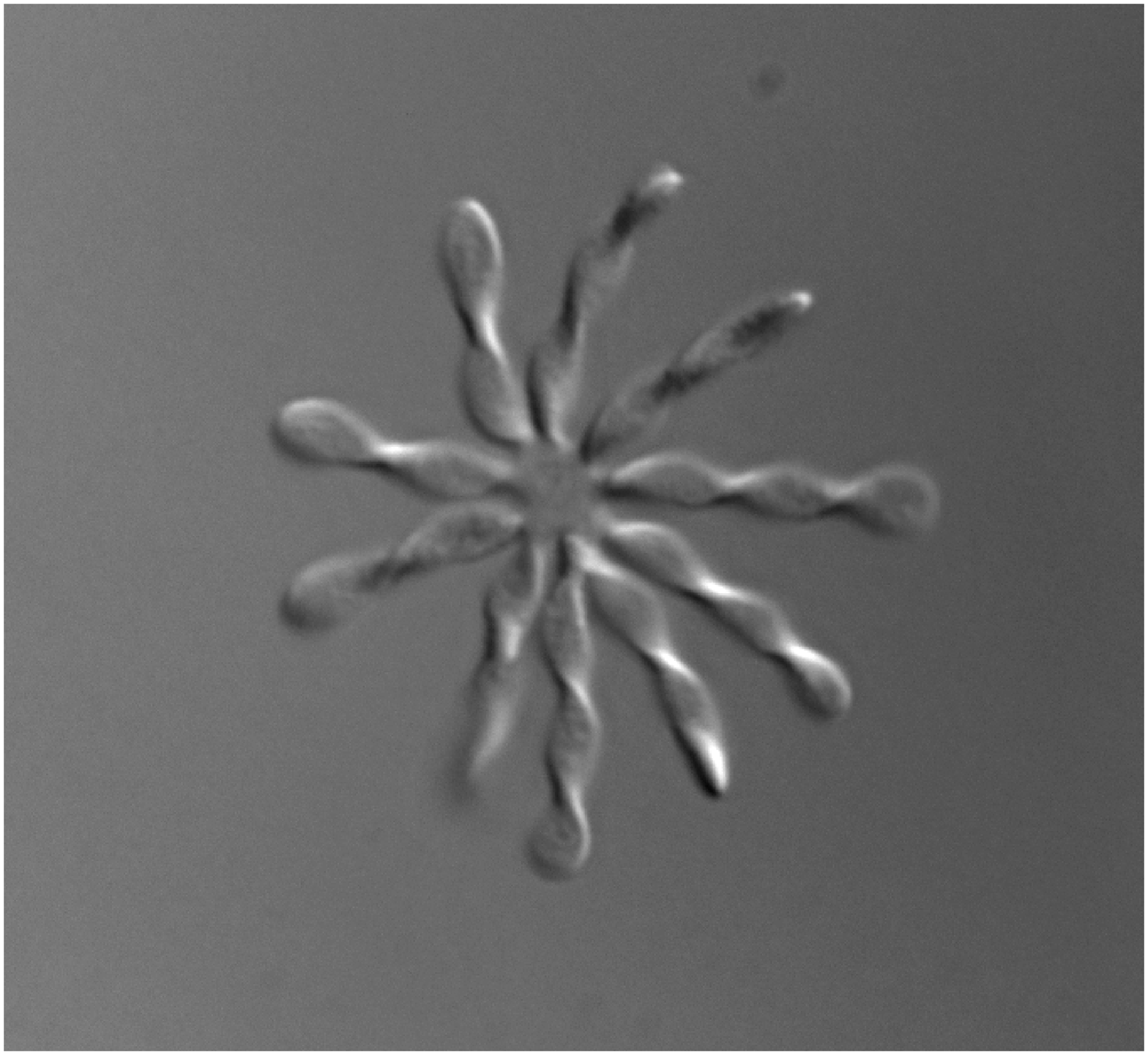}
\caption{Creation of twisted ribbons as temperature is lowered (increasing chirality) \cite{starfish}. At first ripples form at the edge of the membrane (leftmost figure). As the temperature is furthered lowered (going from left to right in the images) the protrusions transform into ribbons as shown.}
\label{Fig:ripple}
\end{figure}

Self-assembled Sm-\textit{A}$^*$ monolayers composed of chiral \textit{fd} viruses in the presence of a polymer depletant have attracted much attention recently \cite{Barry2009,Gibaud2012}.  An \textit{fd} virus is approximately $880 nm$ long and $6.6nm$ in diameter and its chirality can be tuned by temperature. These viruses behave like hard rods in the self-assemblies with a persistence length that is several times the virus length in the temperature range of the experiments \cite{BarryBeller2009}. Upon varying the concentration of the viruses or the polymer depletant, a variety of structures have been observed such as nematic tactoids, twisted ribbons and flat circular disks, with the latter two being Sm-\textit{A}$^*$ monolayer membranes. A first order phase transition between large flat disks and twisted ribbons occurs upon lowering the concentration of the polymer depletant or lowering the temperature (thus increasing the strength of the chirality). During the transition from disks to ribbons, ripples start to form on the edge in the plane of the disk. The ripples grow in the in-plane radial direction as protrusions, making the membrane resemble the shape of a starfish as shown in Fig.~\ref{Fig:ripple}. To the accuracy of the experimental measurement (approximately 0.5 $\mu m$) the membrane remains flat. As the transition proceeds, e.g., by continuing to lower the temperature, the protrusions transform into ribbons and consume nearly all of the material inside the disk. This transition is completely reversible upon increasing the temperature. It is also possible to transform a flat disk into a metastable twisted ribbon using optical traps attached at opposite edges to stretch the membrane \cite{Fig:stretch}. The membrane remains flat during the initial stage of stretching and then abruptly transforms into a twisted ribbon once the extension reaches a critical value as shown in Fig.~\ref{stretch}. The resulting twisted ribbon relaxes back to a flat disk when the traps are removed.

\begin{figure}
\centering
\includegraphics[width=0.5in]{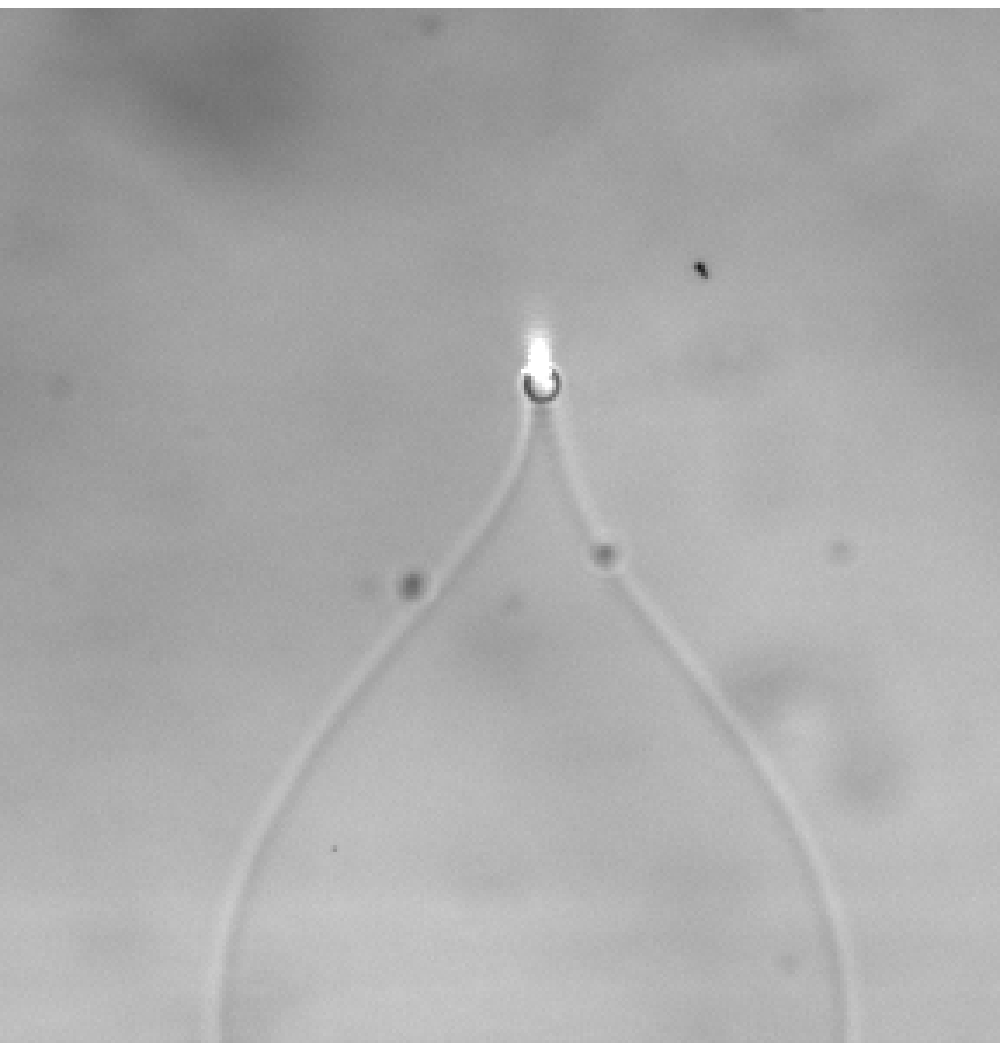}
\includegraphics[width=0.5in]{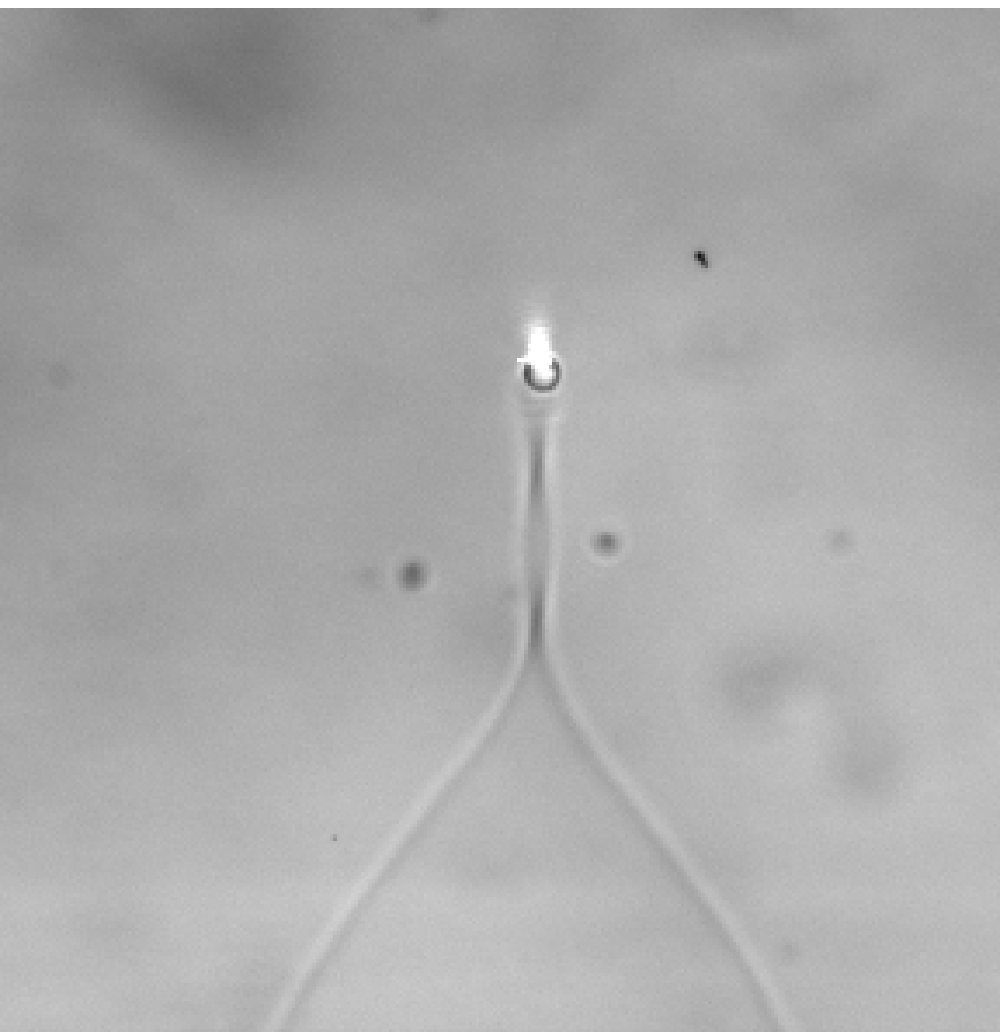}
\includegraphics[width=0.5in]{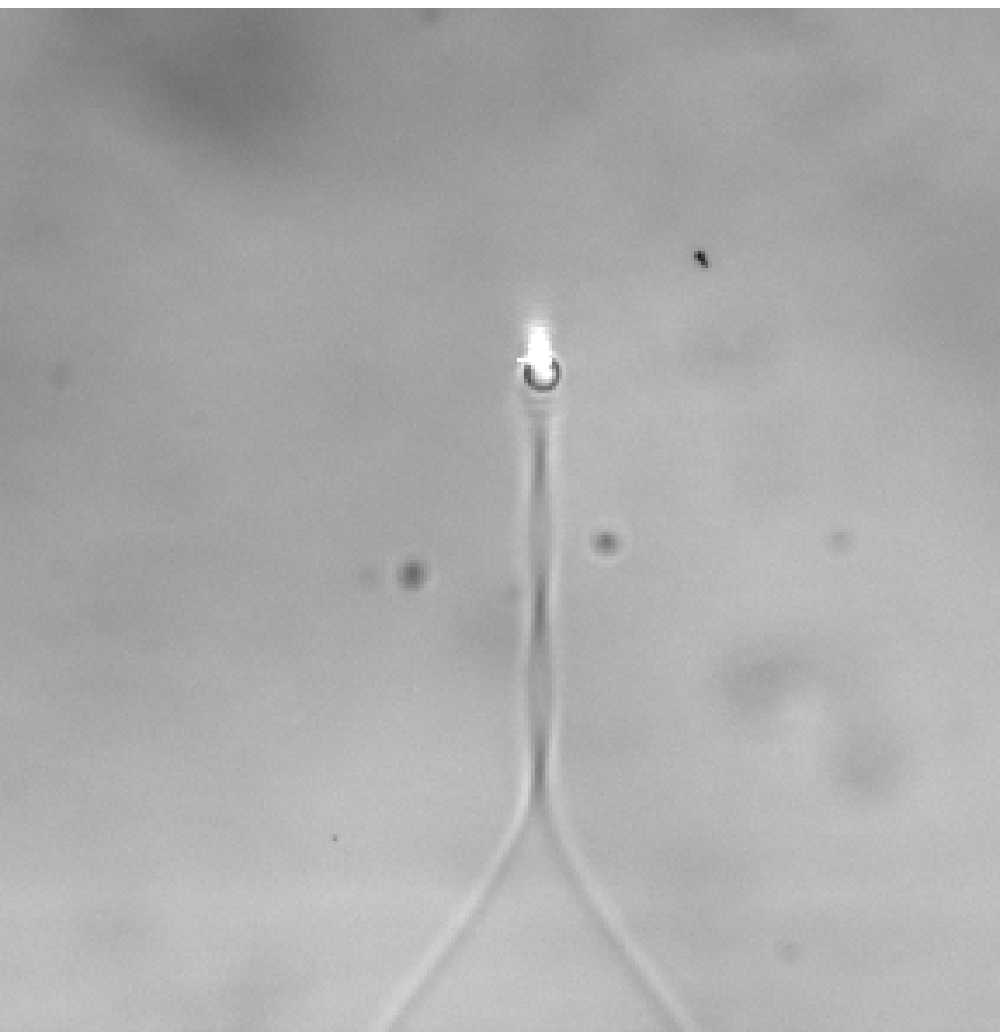}
\includegraphics[width=0.5in]{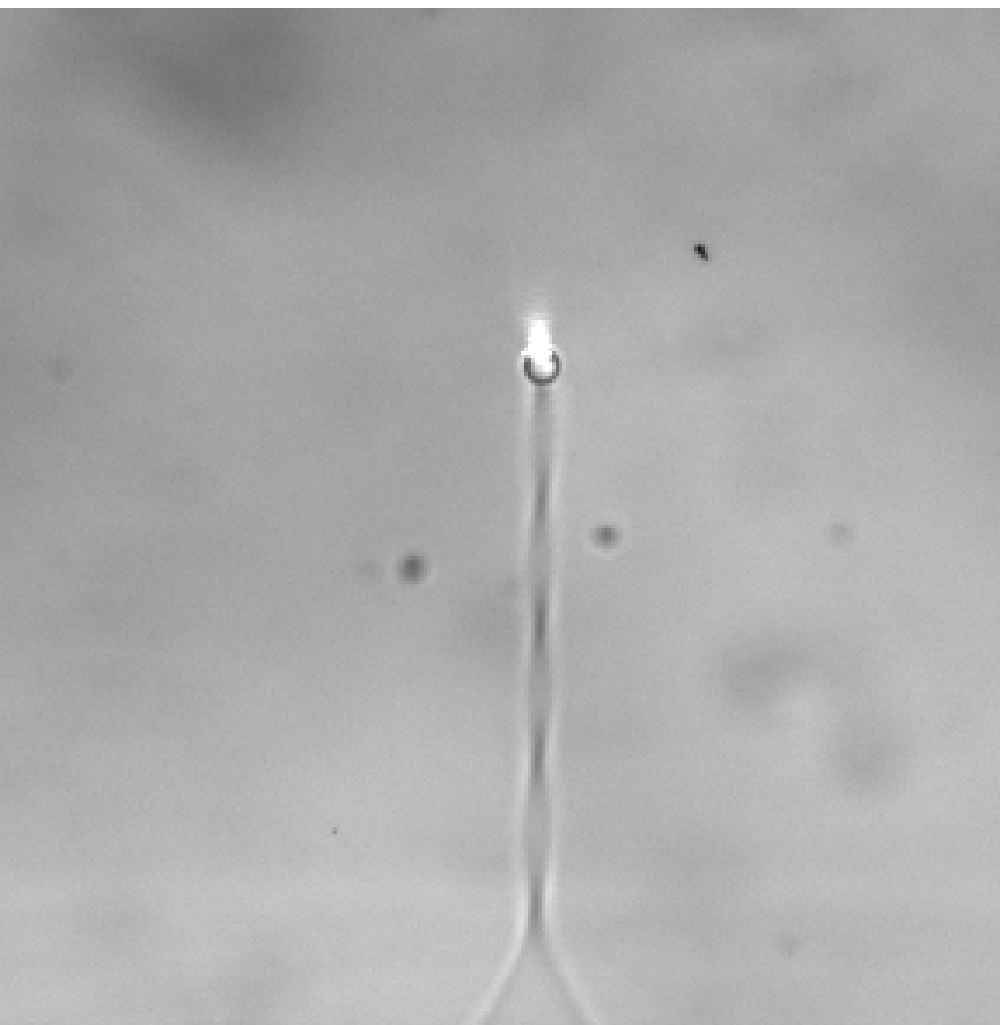}
\caption{Stretching of a membrane by optical traps \cite{stretch}(only one trap, the bright spot, is shown). The membrane is increasingly stretched from one image to the next, from left to right. Once the membrane has been sufficiently stretched a twisted ribbon forms as can be seen in the two rightmost figures.}
\label{Fig:stretch}
\end{figure}

Theories describing the flat disks and twisted ribbons have been constructed \cite{Barry2009,Pelcovits2009,Kaplan2010} using the de Gennes model for the Sm-\textit{A} phase \cite{deGennes1972} generalized to include chirality, and in addition, in the case of the ribbons, the Helfrich model \cite{Helfrich1973,Helfrich1975} for the surface bending energy. These theories use a simple form for the edge energy, the interaction of the rods at the edge with the polymer depletant, proportional to the edge length. A more realistic model incorporating surface tension and the ``melting'' of the smectic order at the edge has recently been developed \cite{KaplanMeyer}. When applied to twisted ribbons the theory with the simple edge model yields good qualitative agreement with experimental measurements of the ribbon's pitch to width ratio providing that the Gaussian curvature modulus appearing in the Helfrich energy is positive, in contrast to the negative values typically measured in lipid monolayers or bilayers \cite{Marsh}. By comparing the free energy per unit area of a twisted ribbon with the corresponding energy of a large, flat membrane, a first-order phase transition between the two structures was found in agreement with experimental observation. However, the predicted value for the edge energy modulus was found to be an order of magnitude lower than that measured experimentally, presumably due to the very simple nature of the edge energy model.

The theory used to study the twisted ribbon is quite general and can be applied to a monolayer of any shape. The free energy $F$ of the monolayer is given by:
\begin{equation}
F=\int(f_H+f_n)\,dA+\gamma\oint dl
\label{equ:FETotal}
\end{equation}
where $f_H$ and $f_n$ are the Helfrich and de Gennes free energy densities respectively and $\gamma$ is the edge energy modulus (``line tension''). The Helfrich free energy density is given by
\begin{equation}
f_H=\frac{1}{2}k(2H)^2+\bar{k}K_G
\label{equ:FHelfrich}
\end{equation}
where $H$ and $K_G$ are the mean and Gaussian curvature of the surface respectively, $k$ is the mean curvature modulus (or ``bending rigidity'') and $\bar{k}$ is the Gaussian curvature modulus. We have assumed a zero spontaneous curvature because of the up--down symmetry of the \textit{fd} viruses.

The de Gennes free energy density in the one-elastic constant approximation with the assumption of perfect smectic order is given by \cite{Kaplan2010}:
\begin{equation}
\label{eq:deG}
\begin{split}
f_n=\frac{1}{2} K&\left[ \left(\nabla\cdot \mathbf n\right)^2-2q\mathbf n\cdot\left(\nabla\times \mathbf n\right)+\left(\nabla\times\mathbf n\right)^2+q^2\right]\\&+\frac{1}{2} C \sin^2{\theta}.\end{split}
\end{equation}
where $\theta$ is the relative tilt angle of the director $\mathbf{n}$ with respect to the local surface normal, $K$ is the single Frank elastic constant, $q$ is magnitude of the spontaneous twist wave vector arising from molecular chirality and $C$ is the tilt free energy modulus. The twist penetration depth is given by $\lambda_t=\sqrt{K/C}$.

A monolayer of general shape can be modeled mathematically as a two-dimensional surface embedded in three dimensions. The surface is given by a position vector $\mathbf{Y}(u_1,u_2)$ parameterized by two coordinates, $u_1$ and $u_2$.
In terms of this position vector the total free energy Eq.~(\ref{equ:FETotal}) is found after some calculation to be \cite{Kaplan2010}:
\begin{widetext}
\begin{align}
F=&\frac{K}{2}\int
\Bigg\lbrace(\partial_jn_j+n_l\Gamma_{jl}^j)^2-2q \frac{\epsilon_{3ji}}{\sqrt{g}}\left[\cos{\theta}\left(g_{ik}\partial_j n_k+g_{il} n_k \Gamma^l_{jk}\right)-\left(2 n_k n_l g_{il} L_{jk}+g_{il} n_l\partial_j\cos{\theta}\right)\right]
\nonumber\\&+\left(\frac{\epsilon_{3ji}}{\sqrt{g}}[(g_{ik}\partial_jn_k
+g_{il}n_k\Gamma_{jk}^l)\hat{\mathbf{N}}-(2n_kL_{jk}+\partial_j\cos\theta)\mathbf{Y}_i]\right)^2\Bigg\rbrace\sqrt{g}
\,du_1du_2\nonumber\\&+\frac{C}{2}\int \sin^2\theta\sqrt{g}\,du_1du_2+\bar{k}\int\frac{L}{g}\sqrt{g}\,du_1du_2
+\frac{k}{2}\int(g^{ij}L_{ij})^2\sqrt{g}\,du_1du_2+\gamma\oint\,dl
\label{equ:FE}
\end{align}
\end{widetext}
where the indices $i,j,k=1,2$ and we sum over repeated indices.   The  tensors $g_{ij}$ and $L_{ij}$ are the first and second fundamental forms of the surface, respectively;  $\Gamma^k_{ij}$ are the Christoffel symbols, $g$ is the determinant of $g_{ij}$, $\epsilon_{ijk}$ is the antisymmetric Levi--Civita tensor, and  $\partial_i\equiv \partial_{u_i}$. The director field $\mathbf{n}$ is expressed in a local basis formed by $\mathbf{Y}_1$, $\mathbf{Y}_2$ and the layer normal $\hat{\mathbf{N}}$. More details on these quantities can be found in Refs.~\cite{Kaplan2010,OuYang1990,OuYang1989}.

Henceforth we use dimensionless units measuring lengths in units of the twist penetration depth $\lambda_t$ and energy in units of the Frank constant $K$. The twist penetration depth has been measured experimentally with a value of 0.5$\mu$m\cite{Barry2009}. The twist Frank constant in bulk \textit{fd} systems has also been measured experimentally \cite{Dogic1} with a concentration dependent value on the order of $10^{-7}$ dynes $\approx 100k_B T$ at room temperature. Thus, thermal fluctuations of the director are negligible, as noted already in Refs.~\cite{Barry2009, Kaplan2010}, where the free energy was minimized and very good agreement was found between the predicted director pattern and experimental measurements.

In this paper we use this model to study the two mechanisms for transforming a flat membrane into a twisted ribbon described above. In Sec.~\ref{ripples} we study a model of the rippled disks observed when the transition to a twisted ribbon is driven by a decrease in  the concentration of polymer depletant which we assume is proportional to the line tension $\gamma$. We find that as $\gamma$ is decreased the size of the ripples in both the radial and out of plane directions grow. In a narrow range of $\gamma$ there is a very rapid increase in the radial size of the protrusions while the height of the ripples remains nearly constant with a value too small to be observed experimentally. In Sec.~\ref{stretching} we consider the transition to a twisted ribbon as induced by an external force, namely, a pair of optical traps which stretch the membrane (one trap holds the membrane in place while the second one pulls the edge of the membrane at the point of attachment). We consider a semi-infinite membrane and assume that the stretching produced by the second trap gives rise to a small protrusion at the edge of the membrane. We then add a ribbon-like perturbation to the tip of protrusion and determine whether the perturbation is energetically favorable. We find that for a large enough protrusion a ribbon-like perturbation is favorable. Due to the complexity of the shape we carry out this analysis for a model which neglects the director field and thus includes only the Helfrich and edge energies.
 We offer concluding remarks in Sec.~\ref{conclusions}.

\section{Instability related to the spontaneous phase transition}
\label{ripples}
\begin{figure}
\centering
\includegraphics[width=3.0in]{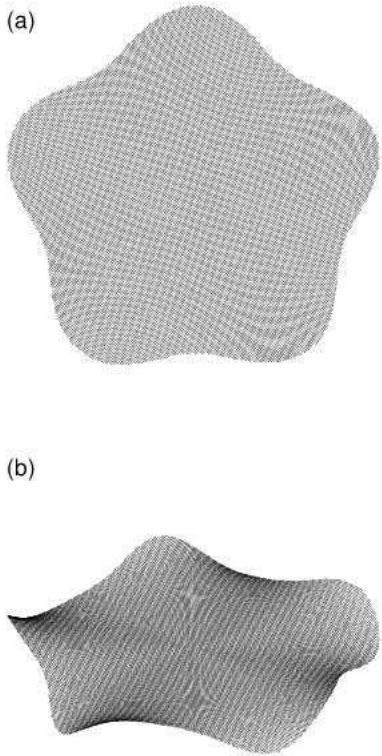}
\caption[Shape of a rippled disk.]{Shape of a rippled disk given by Eqs.~(\ref{equ:ShpRipDisk}) and (\ref{equ:EdgeRipDisk}) with $A_z=0.5$, $A_r=0.5$, $\psi=0.0$, $a=5$, $b=1.0$ and $R_0=5.0$. (a) Viewed from above. (b) Viewed from a tilted angle.}
\label{Fig:ShpRipDisk}
\end{figure}

\begin{figure}
\centering
\includegraphics[width=2.5in]{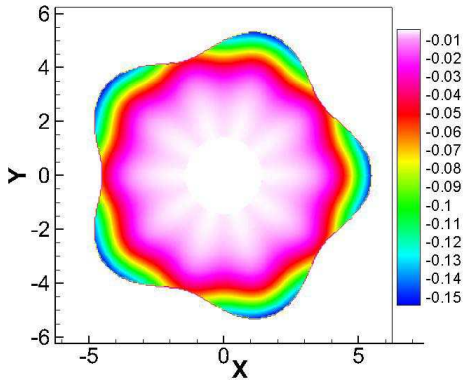}
\caption {Contour plot of the Gaussian curvature for the shape shown in Fig.~\ref{Fig:ShpRipDisk}. Note the maximum negative value at each outward bulge.}
\label{Gaussian}
\end{figure}

We use a model \cite{MeyerDiscussion} of the rippled disks where the ripples are described by sinusoidal waves both in the plane of the disk and out of the plane. While the experiments appear to observe flat membranes this is only to an accuracy of approximately 0.5 $\mu m$ or one unit of dimensionless length. We allow for the possibility of out of plane fluctuations in order to incorporate negative Gaussian curvature in the structure which we believe is necessary for the formation of twisted ribbons. We assume that the out-of-plane height decays exponentially into the interior of the disk so that the ripples are confined to the edge region. The disk lies in the $x\textit{--}y$ plane with its center at the origin. The height of the disk in the $z$ direction is given by:
\begin{equation}
h(r,\phi)=A_z\sin(a\phi)\exp[-b(R_0-r)].
\label{equ:ShpRipDisk}
\end{equation}
The radial coordinate $R(\phi)$ of the edge of the disk is given by:
\begin{equation}
R(\phi)=R_0+A_r\cos(a\phi+\psi).
\label{equ:EdgeRipDisk}
\end{equation}
Here $r$ and $\phi$ are polar coordinates in the $x\textit{--}y$ plane, $A_z$ and $A_r$ are the amplitudes of the ripples in the $z$- and radial directions respectively, $R_0$ is the radius of the disk in the absence of ripples, $a$ is the number of the ripples along the edge, $b$ measures the decay of the height of the ripples into the interior of the disk and $\psi$ is the phase difference between the in-plane ripples and out-of-plane ripples.  The radial coordinate $r$ appearing in the height function $h(r,\phi)$ ranges from a cutoff $r_0 \neq 0$ up to $R(\phi)$. For $r <r_0$ we assume that the disk is perfectly flat. We introduce a nonzero cutoff $r_0$ because in the limit $r_0 \rightarrow 0$ the Gaussian curvature of the above shape diverges due to the factor $\sin(a\phi)$ in the height function. The value of $r_0$ is determined as follows. We calculate the Gaussian curvature $K_G(r_0,\Delta r)$ for a ring of inner radius $r_0$ and outer radius $r_0+\Delta r$ with $\Delta r$ small compared to $R_0$. We first choose $r_0$ of order $R_0$ and then decrease the value of $r_0$ until we reach the point where $K_G(r_0,\Delta r)$ begins to increase. This criterion determines the value of $r_0$ which is of order one for the shapes we consider here. Although the introduction of this limiting radius produces a discontinuity in the shape, the discontinuity is about $1\%$ of $A_z$ and does not affect our physical results.  Figure~\ref{Fig:ShpRipDisk} shows an example of the rippled disk model. In Fig.~\ref{Gaussian} we show a contour plot of the Gaussian curvature of this shape. Note that the radial bulges are the regions of maximum negative value of the Gaussian curvature which we conjecture act as seed points for the growth of twisted ribbons.

Although it is not clear from experiments whether these rippled disks are equilibrium structures, we assume they are and solve for the director field by minimizing the free energy, Eq.~(\ref{equ:FE}), computed for this shape.
With an analytic form of the shape specified, the differential geometry quantities needed in Eq.~(\ref{equ:FE}) can be computed explicitly. In principle we could then follow the approach of Ref.~\cite{Kaplan2010} and derive the corresponding Euler-Lagrange equations for the director field. However because of the complexity of the shape, the resulting equations are impossible to solve explicitly even using numerical solvers. Instead we discretize the underlying $x\text{--}y$ plane using a square lattice of grid size $0.05$ and carry out an MC simulation at low temperature ($10^{-4}$ in dimensionless energy units with $k_B=1$) varying the geometrical parameters $A_z$ and $A_r$ of the shape. We note that our dimensionless energy unit, $K=1$, corresponds to approximately $100 k_B T$ at room temperature using the measured value of the twist elastic constant in $\textit{fd}$ solutions \cite{Dogic1}.

The height of the membrane at each lattice site is given by  Eq.~(\ref{equ:ShpRipDisk}) with Cartesian $x\textit{--}y$ coordinates converted to polar coordinates. The shape of the membrane is fixed and not allowed to vary during the computation. Lattice sites with $r > R(\phi)$ (Eq.~(\ref{equ:EdgeRipDisk})) are excluded except those which have nearest neighbors lying within $R(\phi)$ of the origin. These sites serve as ``ghosts" which allow us to impose a boundary condition on the director field. As in Ref.~\cite{Kaplan2010} we consider free boundary conditions \cite{footnote on fixed BC}, and the directors at the ghost sites are sampled in the MC simulation in the same fashion as the directors inside the disk. A finite difference algorithm was used to compute derivatives of the director field and the height function appearing in Eq.~(\ref{equ:FE}).  In each MC cycle, the director at every lattice site in the disk and at the ghost sites was allowed to rotate with the magnitude of the rotation selected so that the total acceptance ratio was approximately 50\%. We started with a configuration where all the directors are pointing in the $z$ direction and ran 50000 MC cycles to reach equilibrium and then an additional 50000 MC cycles to collect data. Convergence was carefully checked. We have also verified that using a random initial condition for the director field produces the same equilibrium state.  Unlike the usual method in a MC simulation where the free energy at each cycle is recorded and averaged, we recorded the orientations of directors in each cycle, computed an average orientation for each director, and calculated the free energy for this averaged configuration. By doing so, we reduced the effects of thermal fluctuation and the configuration obtained is close to the zero-temperature solution, i.e., the solution to the Euler-Lagrange equations. The validity of this method was checked by comparing the results of this method for a flat circular disk to the results obtained in Ref.~\cite{Pelcovits2009} where the Euler-Lagrange equations were solved explicitly.

We varied $\psi$ in Eq. (\ref{equ:ShpRipDisk}) and found that for viruses with chirality $q>0$  the minimum of the free energy appears at $\psi=0$ as this value makes the handedness of the viruses and the edge of the membrane the same. Viruses with opposite chirality yield a minimum free energy with $\psi = \pi$. Without any loss of generality we consider only the $q>0$ case. In order to study the instability to the formation of twisted ribbons, we vary the amplitude of the ripples, $A_z$ and $A_r$, as the line tension $\gamma$ (a measure of the concentration of polymer depletant) is reduced. We fix $b$ at 1.0 in our dimensionless units. Physically, the length $b$ is determined by the interplay between the deformation in the director field and the bending of the membrane so that one penetration depth is a reasonable choice. We fix the number of ripples $a$ at either 3, 4 or 5. We set the Gaussian curvature modulus $\bar{k}=0.15$ and the chirality $q=0.71$ in accord with the earlier studies of flat disks and twisted ribbons \cite{Barry2009,Pelcovits2009,Kaplan2010}. Because the structure we are studying is not a minimal surface (i.e., one with zero mean curvature) we have allowed for both zero and nonzero values of the mean curvature modulus $k$ as we discuss below.

\begin{figure}
\centering
\includegraphics[width=2.5in]{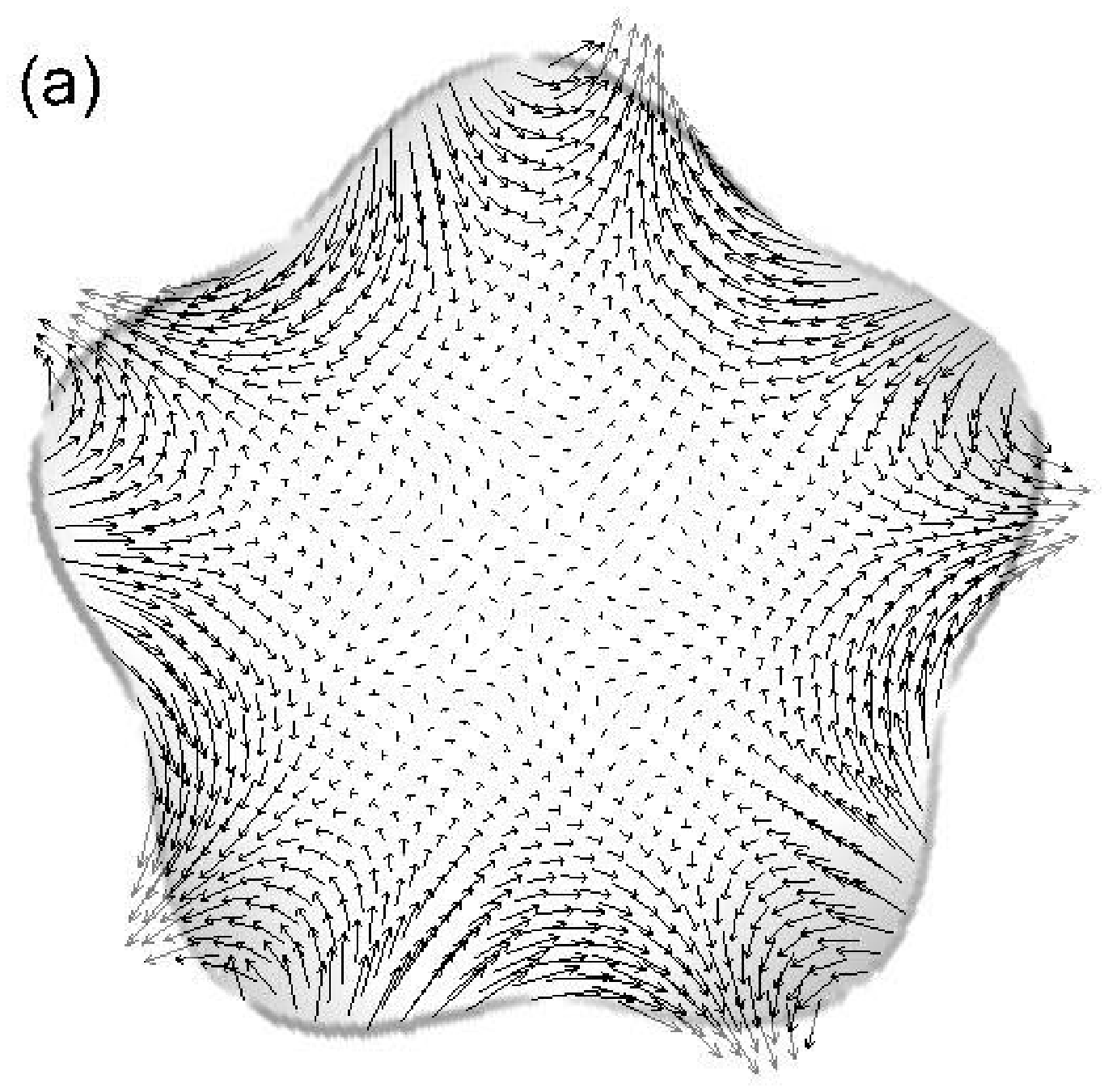}
\includegraphics[width=2.5in]{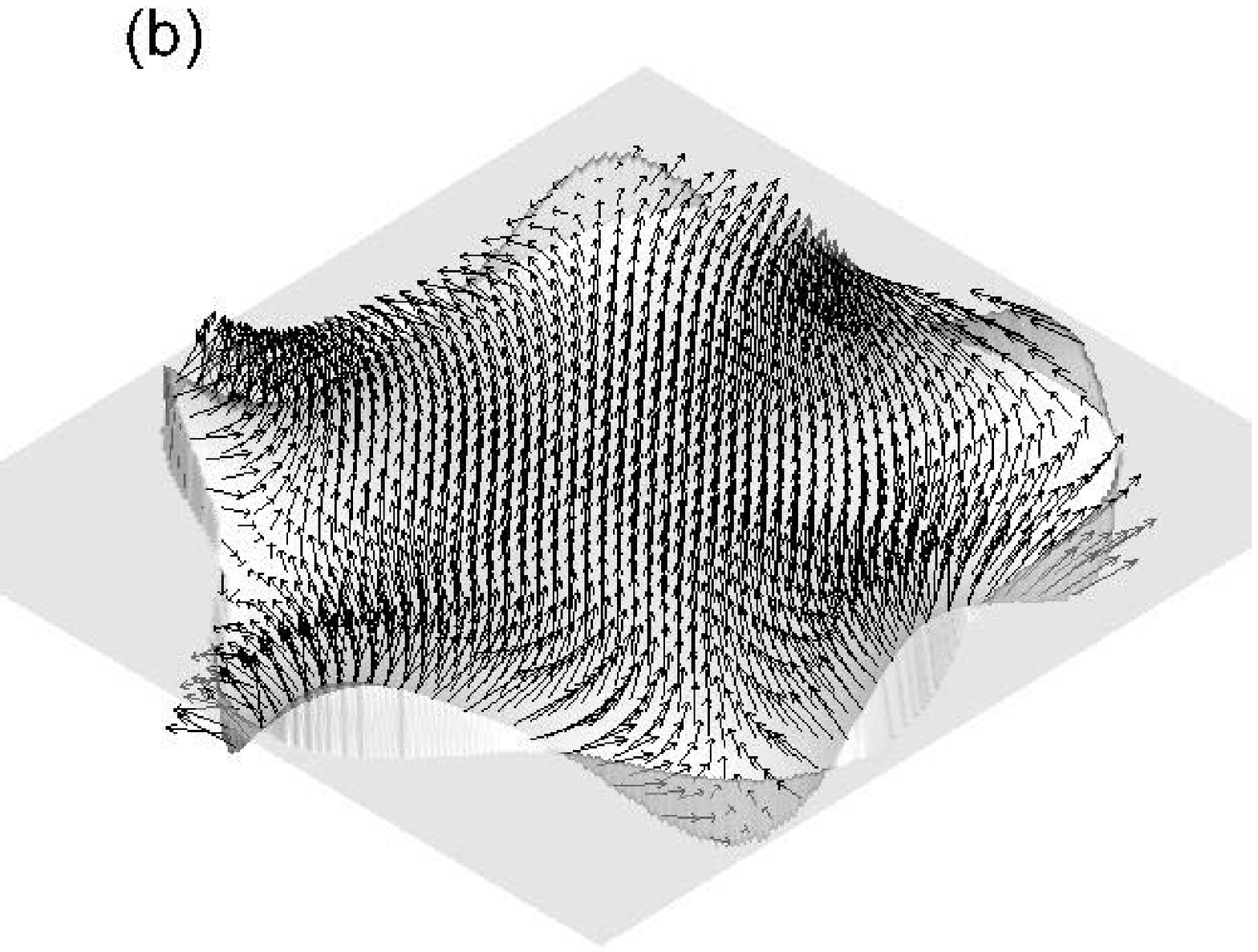}
\caption{Simulation result for the director field on a rippled disk with $A_z=0.5$, $A_r=0.5$, $a=5$, and  $q=0.71$. (a) Viewed from top. (b) Viewed from a tilted angle.}
\label{Fig:SolRipDisk}
\end{figure}
Fig. \ref{Fig:SolRipDisk} shows the simulation result for the director field on a rippled disk with $A_z=0.5$, $A_r=0.5$, $a=5$, and  $q=0.71$. Note that the directors are mostly perpendicular to the $x\textit{--}y$ plane at the center of the disk as expected.

\begin{table}
\centering
\caption{Dependence of the height amplitude $A_z$ and the radial bulge amplitude $A_r$ on the line tension $\gamma$ for a disk with three ripples.  Note that value $A_r=2.2$ in the bottom line is the largest value explored in our simulations yielding a lower bound on the value of $A_r$ that minimizes the free energy. The values here are for the case $k=0, \bar{k} =0.15$. Within each range of $\gamma$ we have varied $\gamma$ with a step size of 0.001.}
\begin{tabular}{|c|c|c|} \hline
$\gamma$ & $A_z$ & $A_r$ \\ \hline\hline
$\ge0.319$ & 0.0 & 0.0 \\ \hline
$0.314-0.318$ & 0.3 & 0.3 \\ \hline
$0.282-0.313$ & 0.3 & 0.4 \\ \hline
$0.274-0.281$ & 0.3 & 0.6 \\ \hline
$0.257-0.273$ & 0.3 & 0.7 \\ \hline
$0.246-0.256$ & 0.3 & 1.0 \\ \hline
$0.244-0.245$ & 0.3 & 1.2 \\ \hline
$0.243$ & 0.3 & 1.3 \\ \hline
$0.224-0.242$ & 0.3 & 1.4 \\ \hline
$0.216-0.223$ & 0.3 & 1.7 \\ \hline
$0.211-0.215$ & 0.3 & 1.8 \\ \hline
$0.200-0.210$ & 0.3 & 2.0 \\ \hline
$0.181-0.199$ & 0.3 & 2.1 \\ \hline
$\le 0.180$ & 0.10 & $\ge2.2$\\ \hline
\end{tabular}
\label{Tab:BehRipDisk}
\end{table}
The results for the values of $A_z$ and $A_r$ that minimize the energy as $\gamma$ is varied are shown in Table \ref{Tab:BehRipDisk} for a disk with three ripples ($a=3$) and $k=0$. Note that the value of $\gamma$ found in Ref.~\cite{Kaplan2010} for the first-order phase boundary between disks and twisted ribbons is 0.276. We have also studied disks with $a=4$ and $a=5$; our results are of the same order of magnitude as those shown in the Table for $a=3$.  The values for $A_z$ and $A_r$  shown were obtained after exploring the range of values: $0<A_z<0.5, 0<A_r<2.2$, a range selected on the basis of the order of magnitude of the experimental measurements.  From Table \ref{Tab:BehRipDisk}, we see that when the line tension is large (well within the disk region of the phase diagram \cite{Kaplan2010}), the free energy minimum corresponds to a disk where both the height and radial size of the ripples vanish. As $\gamma$ is reduced the height and radial size of the ripples both grow. The height $A_z$ grows to a small value of order 0.3, which is well below the experimental resolution for out of plane fluctuations, and thus consistent with experimental observations of ``flat'' rippled disks.  On the other hand, the radial size $A_r$ grows more rapidly inside a narrow range of $\gamma$. Specifically, $A_r$ quadruples from 0.3 to 1.2 in the range between $\gamma\sim0.31$ and $\gamma\sim0.24$, and continues to grow as $\gamma$ is reduced. These results are in qualitative accord with the experimental observations discussed in Sec.~\ref{intro}, namely, that ripples in the radial direction are observed and grow rapidly as the instability to twisted ribbons is approached. Recalling from Fig.~\ref{Gaussian} that ripples with nonzero $A_z$ have negative Gaussian curvature we conjecture that they are the seed points for the growth of twisted ribbons.

The above results are for the case $k=0$. Twisted ribbons and flat disks are minimal surfaces with zero mean curvature; thus, the value of $k$ is irrelevant to an analysis of their energy. However, the present model of a rippled disk is not a minimal surface and thus we consider the effect of a nonzero value of $k$ on our results. We find that for $k \lesssim 0.2$ our results are qualitatively unchanged, however, the critical value of $\gamma$ where the rapid increase of $A_r$ begins is lower. E.g., if $k=0.2$ we find that $A_r$ reaches 2.2 (the largest value we have studied) at $\gamma \sim 0.175$ compared to $\gamma\sim 0.180$ when $k=0$. For $k \gtrsim 0.2$ we find that while the ripples grow in the radial direction as for the larger values of $k$, the membrane remains flat, i.e., $A_z =0$ and thus we presume that these ripples will not form twisted ribbons because of the absence of seed points with nonzero Gaussian curvature. It appears then that the mean curvature energy acts as a barrier between the flat disk and ribbon states; a similar effect will be shown in the next section where we consider ribbon formation on a stretched flat membrane.

\section{Instability upon stretching}
\label{stretching}

In this section we model the instability of a flat membrane that is stretched using optical traps as described in Sec.~\ref{intro}. Experimentally, this instability occurs within the region of the phase diagram where flat disks are the equilibrium shape. We consider large disks and use a semi-infinite model which has been shown \cite{Barry2009} to accurately describe disks of the order of $10\mu m$ in diameter.  We assume that the optical tweezers produce a  protrusion on the edge of the semi-infinite membrane and then determine whether a ribbon-like perturbation attached to the tip of the protrusion is energetically favorable. Given the complexity of the analysis, we neglect the director field and consider only the Helfrich and edge energies.

\begin{figure}
\centering
\includegraphics[width=1.6in]{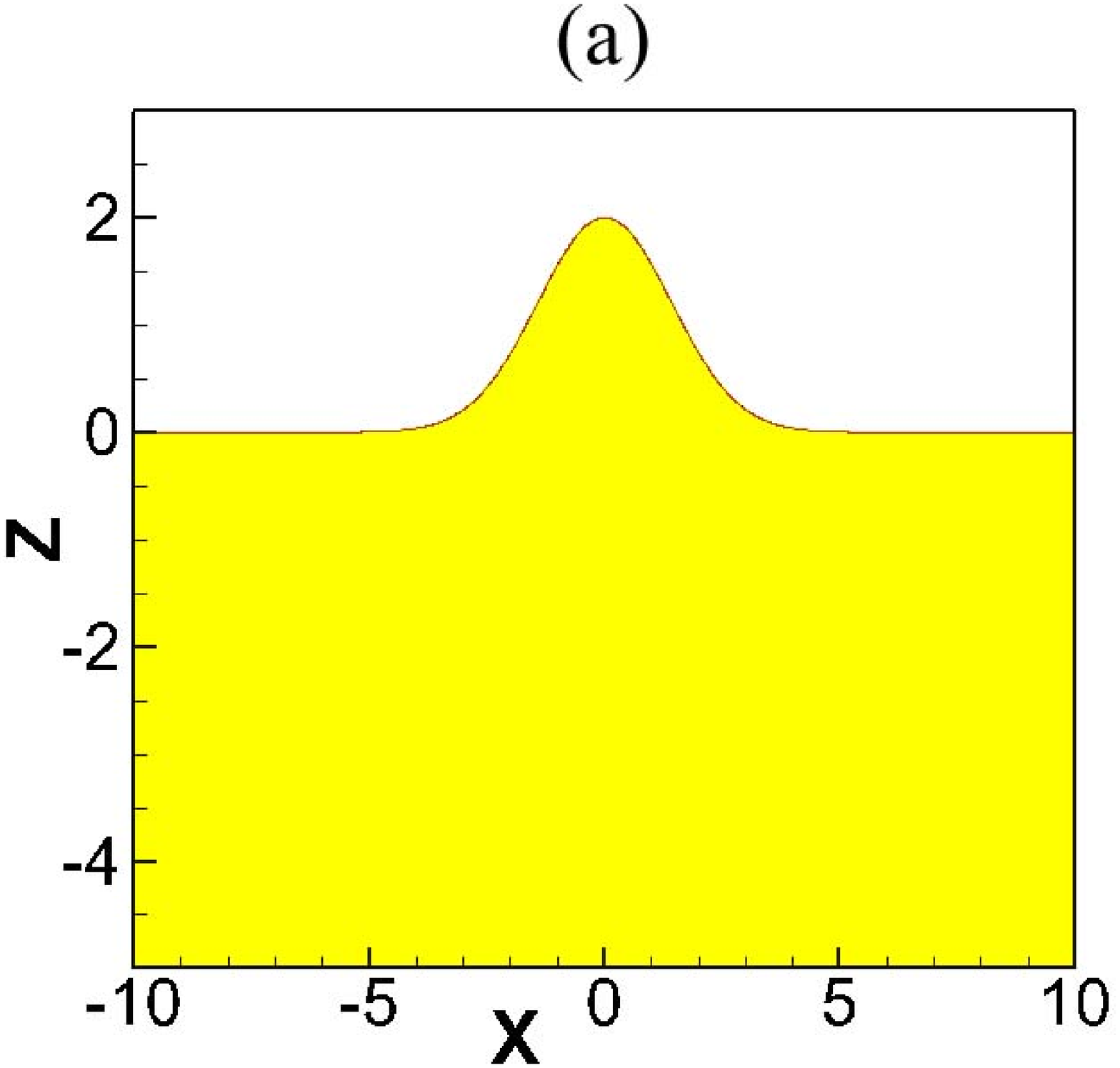}
\includegraphics[width=1.6in]{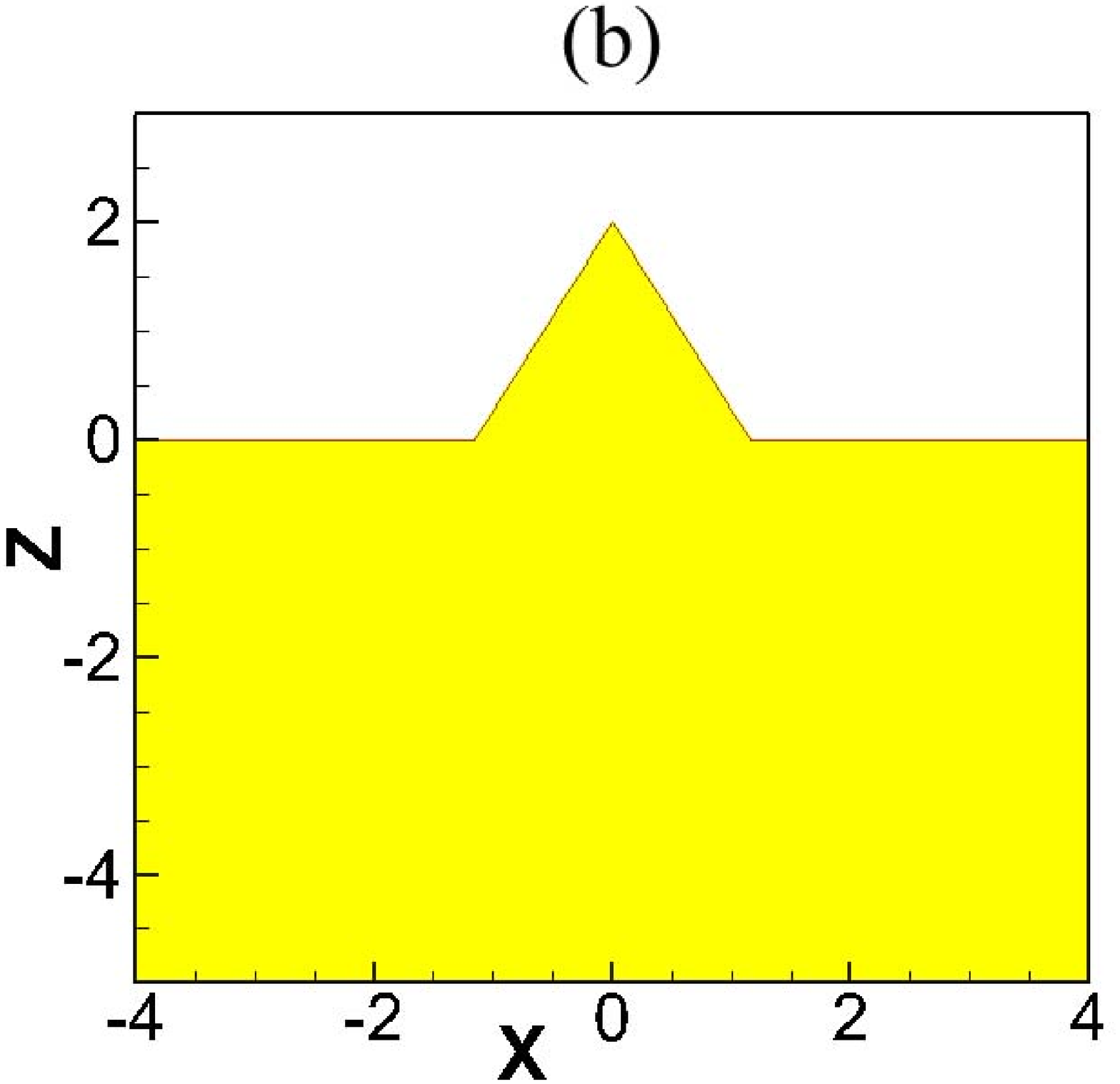}
\caption{Model shapes of a membrane stretched by an optical trap before the ribbon-like perturbation is added. The colored region is the membrane.  The Gaussian bump model is shown in (a) and the equilateral triangle model in (b). These figures correspond to $z_0=2, \lambda =1$}
\label{Fig:Protrusions}
\end{figure}

We assume the semi-infinite membrane lies in the $x\textit{--}z$ half-plane, $z<0$. We consider two different shapes for the protrusion created at $x=z=0$, as shown in Fig.~\ref{Fig:Protrusions}. The first, Fig.~\ref{Fig:Protrusions}(a), is a Gaussian bump given by $z=z_0\exp[-(x/\lambda z_0)^2]$ where $\lambda$ is a dimensionless constant setting the scale of the half width of the Gaussian bump. The second shape for the protrusion is an equilateral triangle with height $z_0$ as shown in Fig.~\ref{Fig:Protrusions}(b).

We denote the position vector of the stretched flat membrane (i.e., either of the shapes shown in Fig.~\ref{Fig:Protrusions}) by $\mathbf{Y}_0(x,z)$. A small ribbon-like perturbation attached to the tip of the protrusion with a pitch axis along the $z$ direction will produce a distortion of the surface in the direction of the surface normal $\hat{\mathbf{N}}$ which lies parallel to the $y$ axis. The position vector of the perturbed surface is then given by the following form:
\begin{equation}
\mathbf{Y}'=\mathbf{Y}_0+\Psi(x,z)\hat{\mathbf{y}}
\label{equ:PerSur}
\end{equation}
where $\Psi(x,z)$ is the amplitude of the perturbation assumed to be small. Using the methods of Ref.~\cite{OuYang1989} we find the  change $\delta F_H$ in the Helfrich free energy due to the perturbation:
\begin{equation}
\label{eq:deltaF}
\delta F_H=2kA+\bar{k}B+O(\Psi^3),
\end{equation}
where
\begin{eqnarray}
A &=& \int(\delta H)^2dxdz=\frac{1}{4}\int(\Psi_{xx}+\Psi_{zz})^2dxdz\\
B &=&\int(\delta K_G)dxdz\int(\Psi_{xx}\Psi_{zz}-\Psi_{xz}^2)dxdz.
\label{equ:PerFHFin}
\end{eqnarray}
The subscripts on $\Psi$ denote partial derivatives taken with respect to the corresponding coordinate. The quantities $\delta H$ and $\delta K_G$ denote the changes in the mean and Gaussian curvatures respectively due to the ribbon-like perturbation. In deriving Eq.~\ref{eq:deltaF} we have  used the fact that the unperturbed shape specified by $\mathbf{Y}_0(x,z)$ is a flat  membrane with $H=K_G=0$.

\begin{figure}
\centering
\includegraphics[width=2.5in]{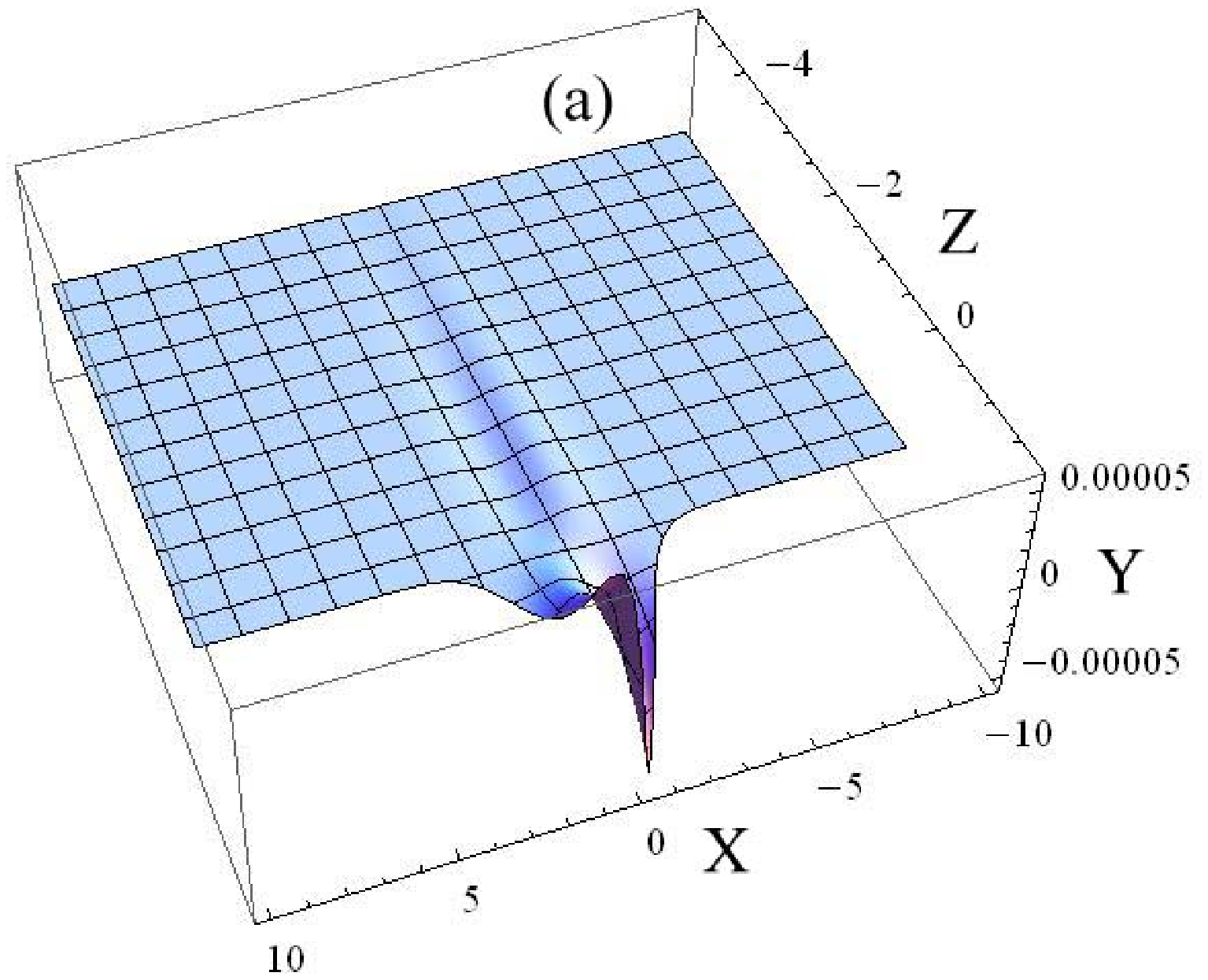}
\includegraphics[width=2.5in]{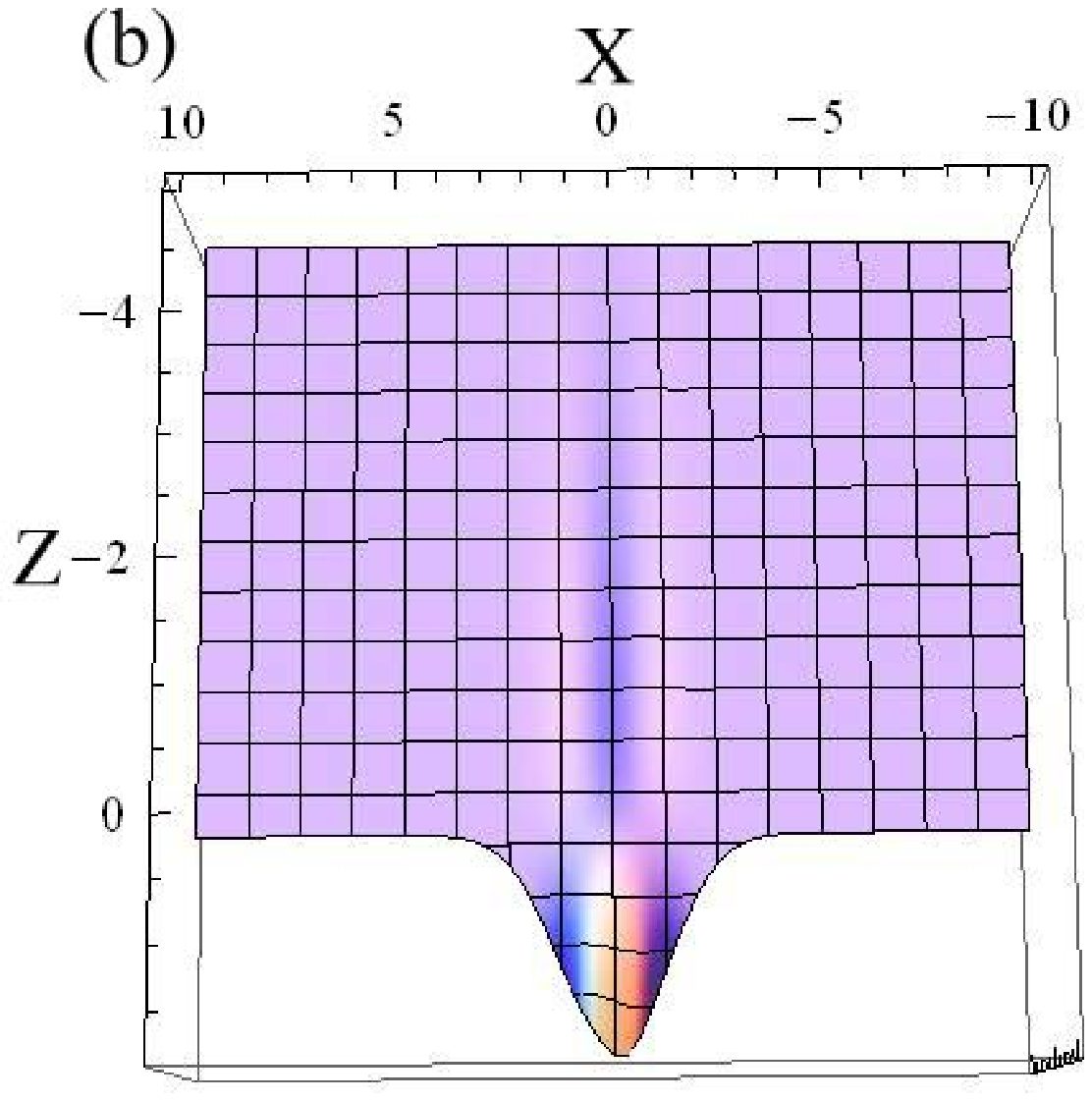}
\caption{Two views of an example of a ribbon-like perturbation Eq.~(\ref{equ:PerFin}) on a flat semi-infinite membrane with a Gaussian bump protrusion on its edge. The parameters are $\lambda_1=\lambda_2=\lambda=1.0$, $b=10^4$ and $z_0=1.778$. Note the scale of the $z$ axis in (a), consistent with the small value of $\Psi$. The view in (b) is along the $y$ axis. Note that the projection of the structure on the $x\textit{--}z$ plane has the same shape as the Gaussian bump, Fig.~\ref{Fig:Protrusions}(a), consistent with Eq.~(\ref{equ:PerSur}). }
\label{Fig:ShpPul}
\end{figure}

For a twisted ribbon with a pitch axis along the $z$ direction and a pitch of magnitude $2\pi|b|$,  the $y$ component of the ribbon's position vector is given by $\Psi=x\tan(z/b)$ \cite{OuYang1990}.  However, because the perturbation occurs only locally around $(0,z_0)$ where the optical trap functions, we assume an exponential decay of the shape in the $x$ direction and in the $z<0$ direction (into the interior of the membrane). Thus, our final expression for the perturbation is given by:
\begin{equation}
\Psi(x,z)=x\tan\left[\frac{ze^{(z-z_0)/\lambda_1}}{b}\right]\exp\left[-\left(\frac{x}{\lambda_2}\right)^2\right]
\label{equ:PerFin}
\end{equation}
where $\lambda_1$ and $\lambda_2$ are parameters defining the rate of decay in the $z$ and  $x$ directions respectively. The parameter $b$ is kept large to ensure that $\Psi$ is small. Fig. \ref{Fig:ShpPul} shows an example of the perturbation on a Gaussian bump protrusion.

Substituting Eq.~(\ref{equ:PerFin})  in Eqs.~(\ref{eq:deltaF})-(\ref{equ:PerFHFin}) yields the change in the Helfrich free energy, $\delta F_H$. The change in  the total free energy (ignoring the director energy) is given by the sum of $\delta F_H$ and the change in the edge energy which is $\gamma$ times the change in the edge length. We evaluated the integrals in Eq.~(\ref{eq:deltaF})  and the change in edge length numerically.  We assume that  $\lambda_1=\lambda_2=1$ using the following reasoning. If these lengths were small compared to the penetration depth, then the surface would be highly curved and there would be a large director energy penalty. Conversely, if these lengths were large, then the perturbation would not be localized at the point on the membrane where the optical tweezer is active. We have verified that a small change in the values of $\lambda_1$ and $\lambda_2$  does not influence our qualitative results. Specifically, we have varied $\lambda_1$ from 0.1 to 1.5, and $\lambda_2$ from 0.6 to 4.0. We also assume that $\lambda=1$, as it is natural to expect that the half width of the Gaussian bump induced by the stretching is of the order of the extent of the stretching, $z_0$. We have verified that changing the value of $\lambda$ from 0.1 to 10 does not influence our qualitative results. Finally, we assume that  $b=10^3$ and have verified that our results are to the accuracy of our calculation quantitatively identical as $b$ is varied from $10^2$ to $10^4$.

\begin{figure}
\centering
\includegraphics[width=2.5in]{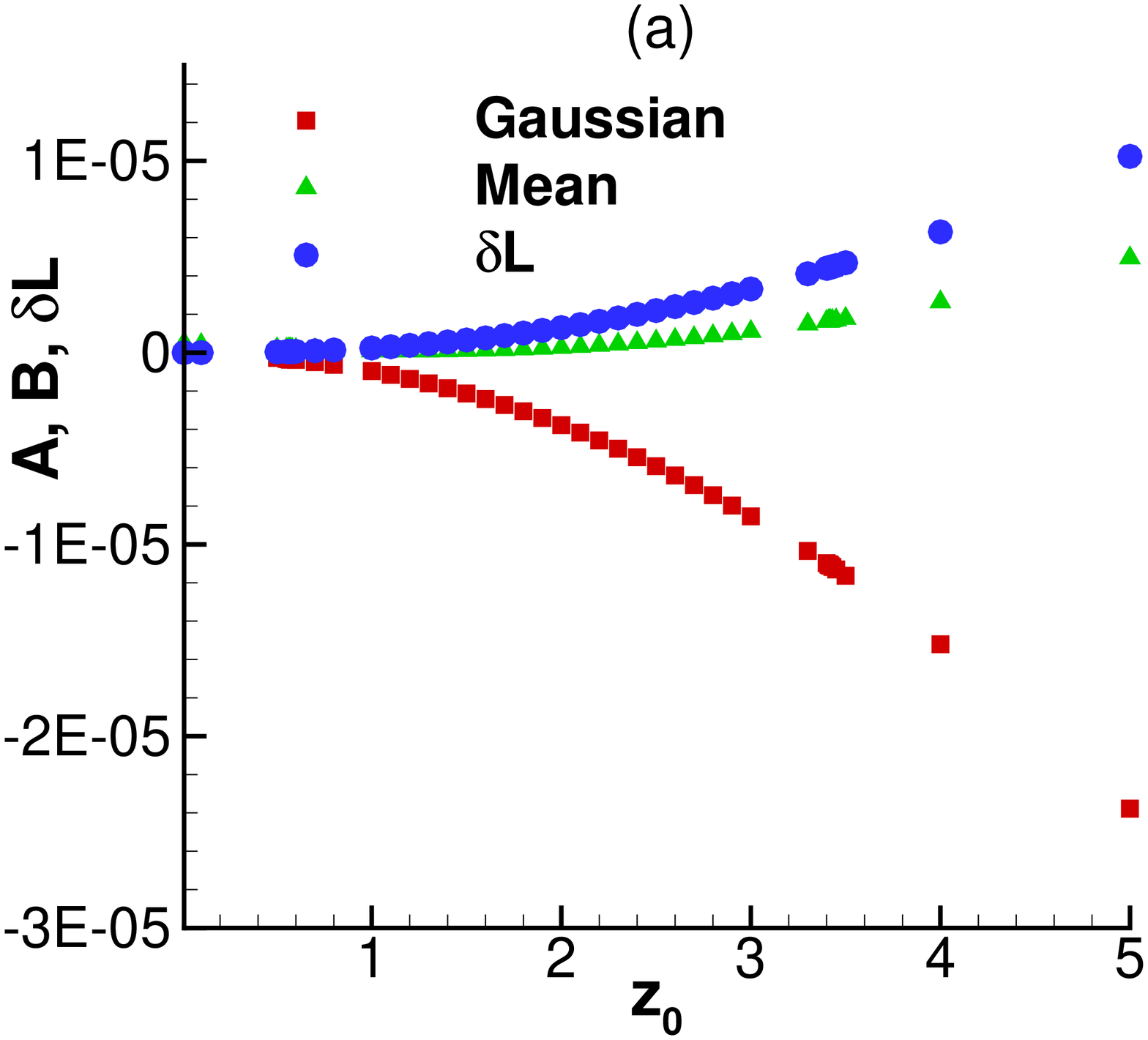}
\includegraphics[width=2.5in]{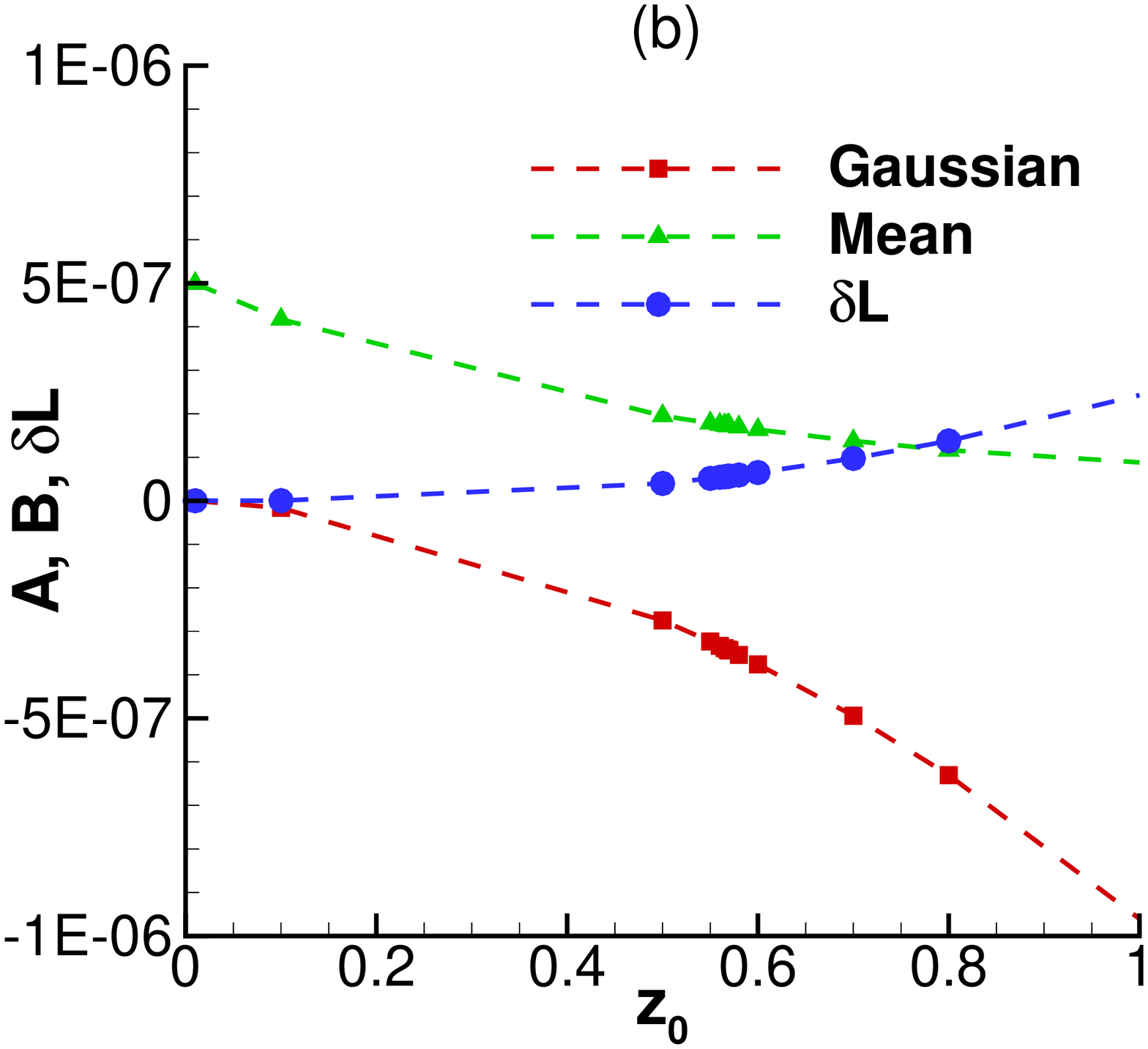}
\caption{(Color online) (a) Integrated Gaussian curvature $B$ (Eq. (\ref{equ:PerFHFin}), squares), integrated mean curvature squared $A$ (Eq. (\ref{equ:PerFHFin}), triangles) and the change in the edge length $\delta L$ (circles) as functions of the extent of stretching $z_0$ for a ribbon-like perturbation (Eq.~(\ref{equ:PerFin})) on a Gaussian bump protrusion. (b) Same as (a) but restricted to the range $0<z_0<1$.}
\label{Fig:TermPul}
\end{figure}

\begin{figure}
\centering
\includegraphics[width=2.5in]{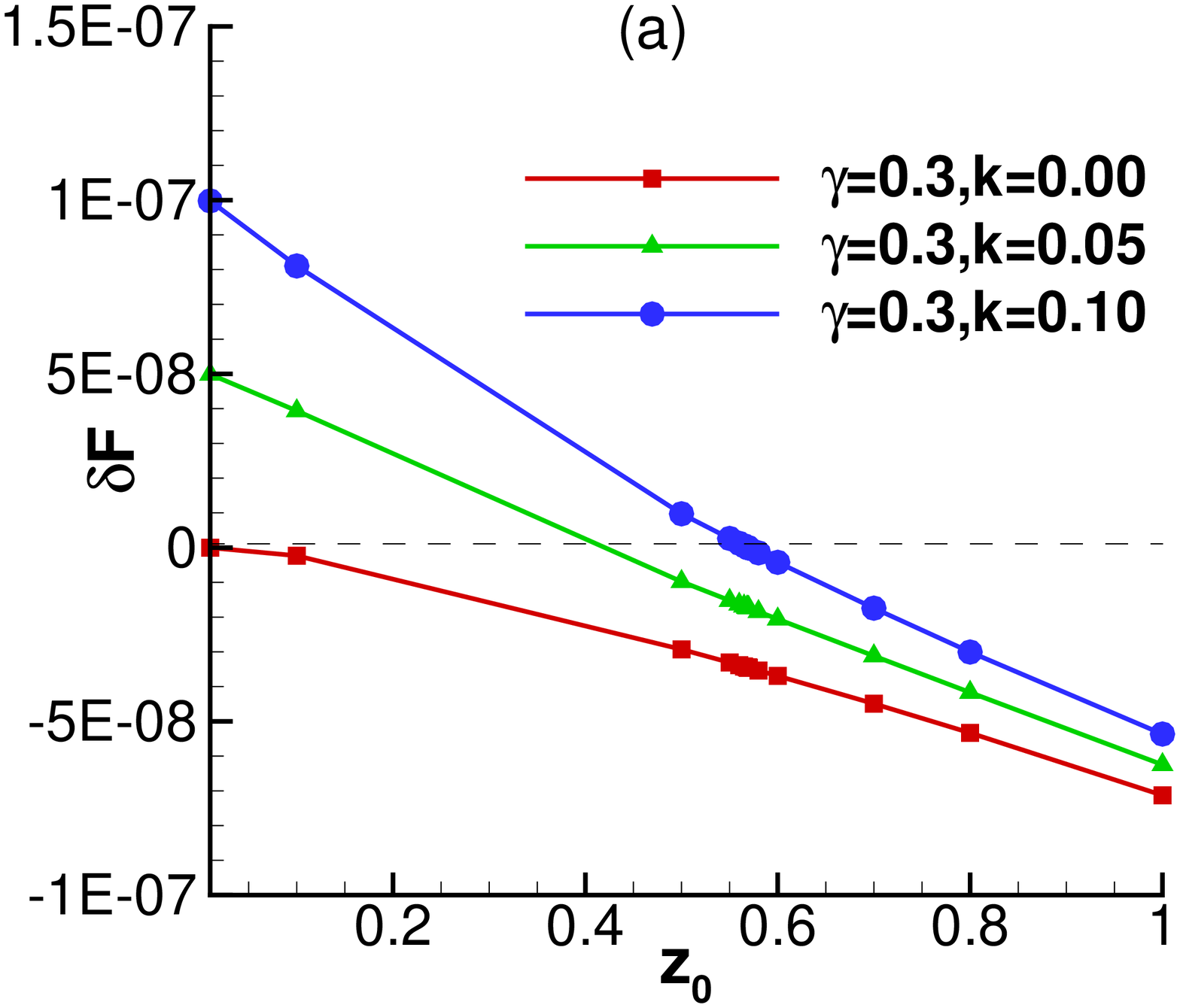}
\includegraphics[width=2.5in]{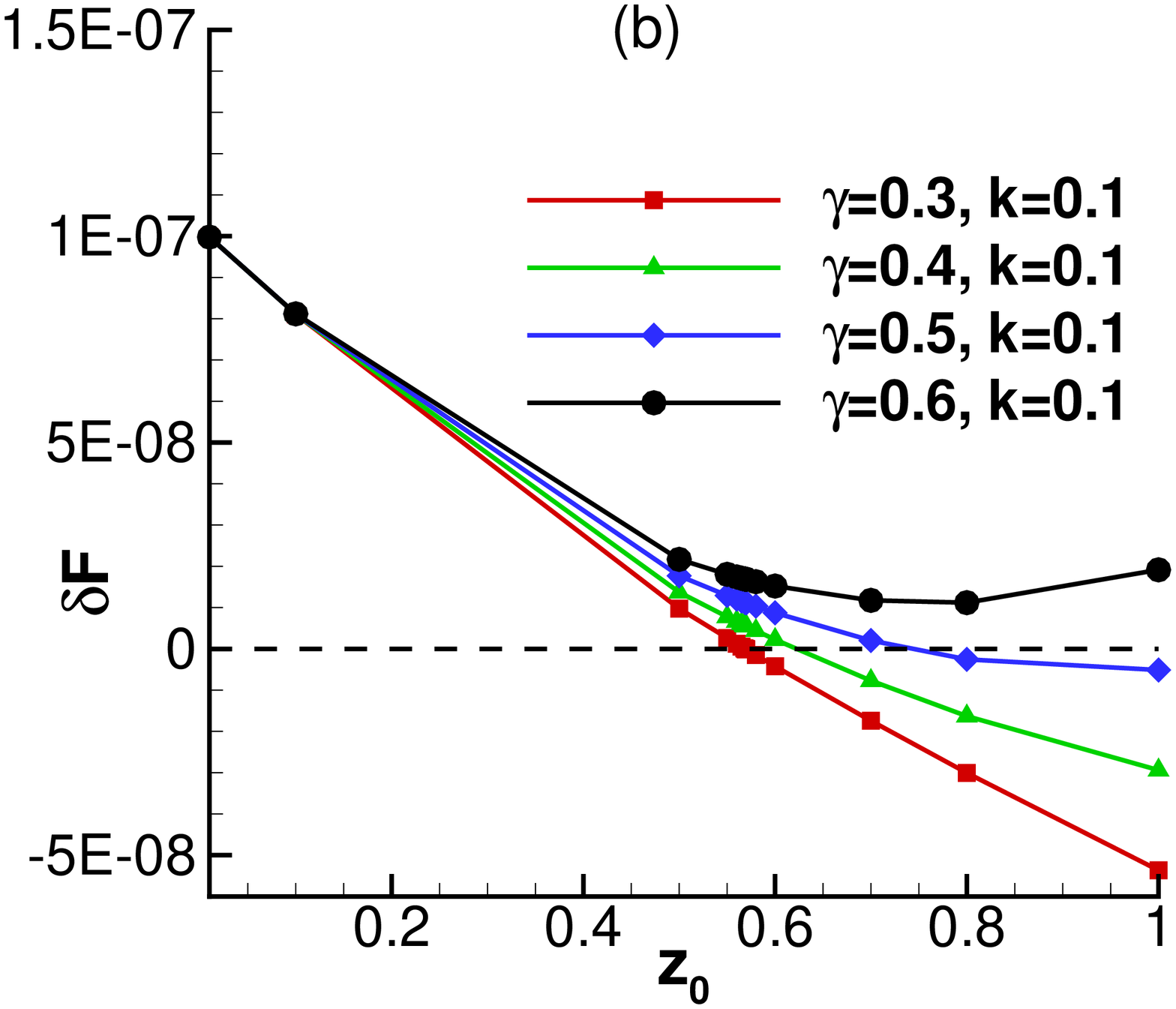}
\caption{The change $\delta F$ in the total free energy (Helfrich plus edge energies) caused by a ribbon-like perturbation on the Gaussian bump as a function of the extent of stretching $z_0$ for different values of the mean curvature modulus $k$ and the line tension $\gamma$. (a) $\gamma=0.3$, $k=0.0,0.05$ and 0.10 from bottom to top. (b) $k=0.10$, $\gamma=0.3,0.4,0.5$ and 0.6 from bottom to top. The dashed horizontal line marks the line at $\delta F=0$ below which the perturbation is energetically favored.}
\label{Fig:BehPul}
\end{figure}

We set the Gaussian curvature modulus  $\bar{k}=0.15$, the same value used in studying the phase diagram of flat membranes and twisted ribbons \cite{Kaplan2010} and considered several values of the mean curvature modulus $k$ and line tension $\gamma$.   Fig.~\ref{Fig:BehPul} shows our results for the change $\delta F$ in the total energy (Helfrich plus edge) as a function of the extent of stretching $z_0$. In Fig.~\ref{Fig:BehPul}(a), $\gamma=0.3$, which is in the flat disk phase of Ref.~\cite{Kaplan2010}, and $k=0.0,0.05$ and 0.10 from bottom to top. In Fig. \ref{Fig:BehPul}(b), $k=0.10$ and $\gamma=0.3,0.4,0.5$ and 0.6 from bottom to top (all in the flat disk phase). The dashed horizontal lines in both figures mark the $\delta F=0$ lines below which the perturbation is favored.  For $k\neq 0$ we see that the perturbation is not favored at the beginning of the stretching process, but once a critical stretching $z_{0c}$ is achieved where $\delta F=0$, the perturbation becomes energetically favorable.  However, the value $z_{0c}$ found in our model is one order of magnitude lower than the value of several penetration depths found in experiment. One possible source of the discrepancy is our neglect of the director energy. Recall that the stretching experiments are performed in the region of the phase diagram where twisted ribbons are not energetically preferred in the absence of an external force. The theory \cite{Kaplan2010} used to study the transition from flat membranes to twisted ribbons induced by lowering the concentration of polymer depletant showed that it is the chiral director energy that drives the transition. Thus, it seems reasonable to expect that including the director energy in the present case should increase the value of the critical stretching. This conclusion is further substantiated by noting that $z_{0c}$ increases with increasing line tension $\gamma$, i.e., increasing the concentration of polymer depletant, making the disks even more energetically favorable compared to twisted ribbons.

As can be seen from Fig.~\ref{Fig:BehPul} (a), when $k=0$  the perturbation is favored at the very beginning of the stretching process, i.e., $z_{0c}=0$. Although  disks and twisted ribbons are both minimal surfaces with zero mean curvature where the value of $k$ is irrelevant to the energetics, our present results suggest that the effect of $k$ is to create a free energy barrier between disks and ribbons. This conclusion is supported by the data shown in Fig.~\ref{Fig:TermPul} where we plot the dependence of  the integrated Gaussian curvature, $B$; the integrated mean curvature squared, $A$; and the change in the edge length, $\delta L$, as functions of the extent of the stretching $z_0$. Note that the integrated mean curvature squared term $A$ remains nonzero (and positive) when the stretching is vanishingly small while the other terms (integrated Gaussian curvature and change in edge length) approach zero. Thus, the ribbon-like perturbation is not favored at the beginning of the stretching process because of the cost in mean curvature energy. While Fig.~\ref{Fig:BehPul} (a) indicates that $z_{0c}$ grows as $k$ is increased in value we find that there is a limit to this growth. For $k\sim 0.6$ we find $z_{0c}\sim1.3$. However, for larger values of $k$ the ribbon-like perturbation is no longer favored no matter how large we make $z_{0c}$.

\begin{figure}
\centering
\includegraphics[width=2.5in]{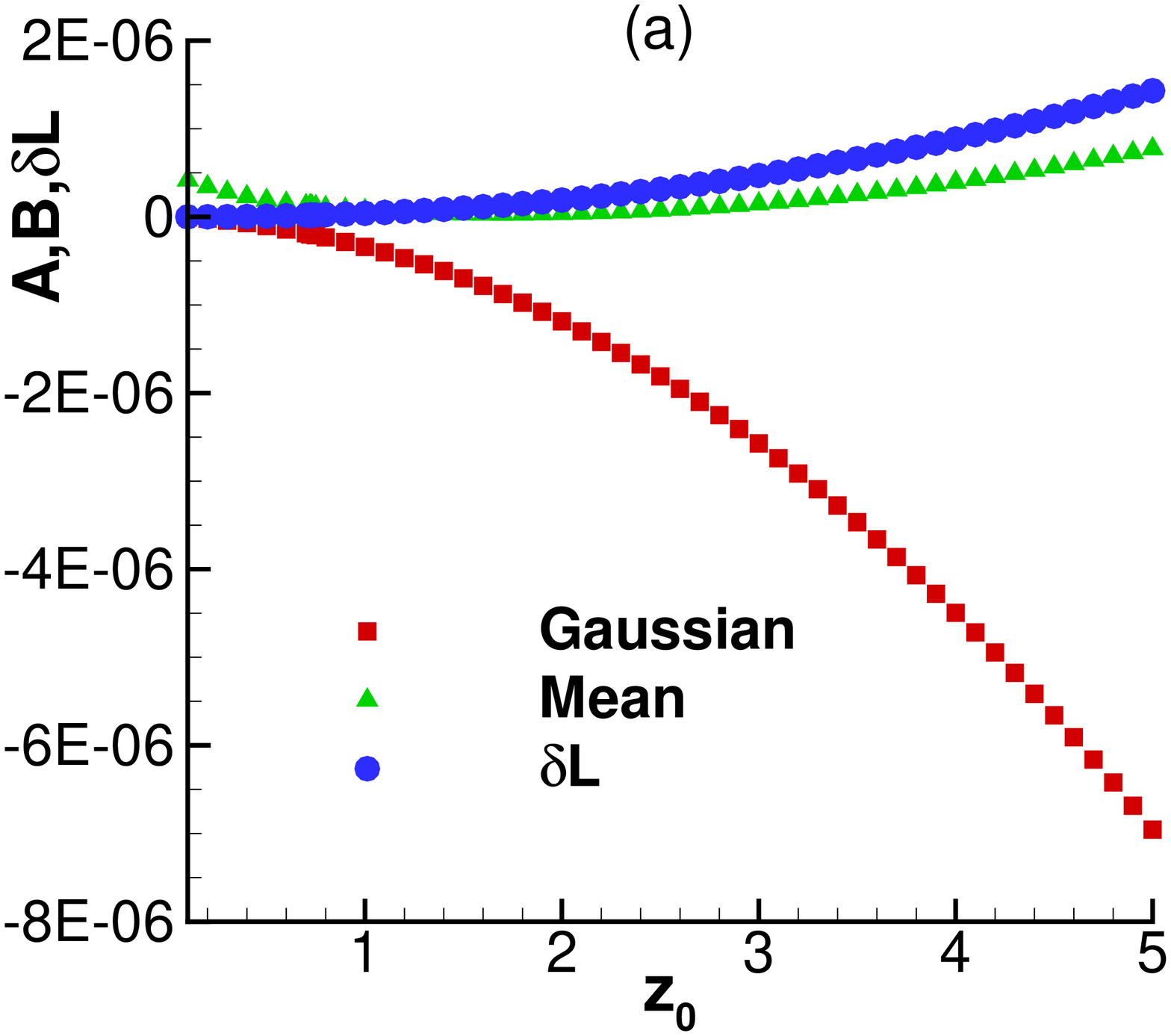}
\includegraphics[width=2.5in]{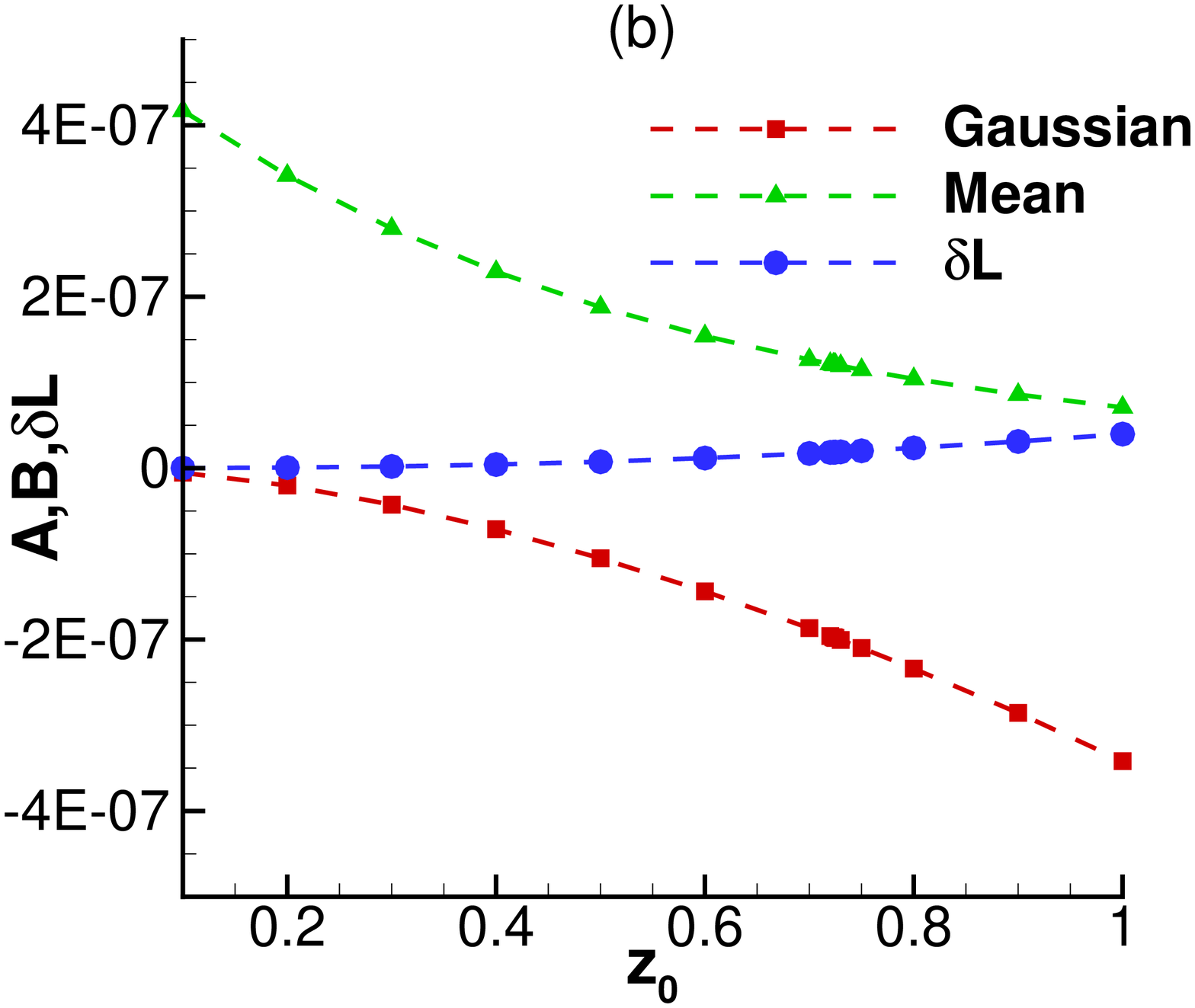}
\caption{(Color online)(a) Integrated Gaussian curvature $B$ (Eq. (\ref{equ:PerFHFin}), squares), integrated mean curvature squared $A$ (Eq. (\ref{equ:PerFHFin}), triangles) and the change in the edge length $\delta L$ (circles) as functions of the extent of stretching $z_0$ for a ribbon-like perturbation (Eq.~(\ref{equ:PerFin})) on a protrusion with shape of a equilateral triangle. (b) Same as (a) but restricted to the range $0<z_0<1$.}
\label{Fig:TermPulTri}
\end{figure}
\begin{figure}
\centering
\includegraphics[width=2.5in]{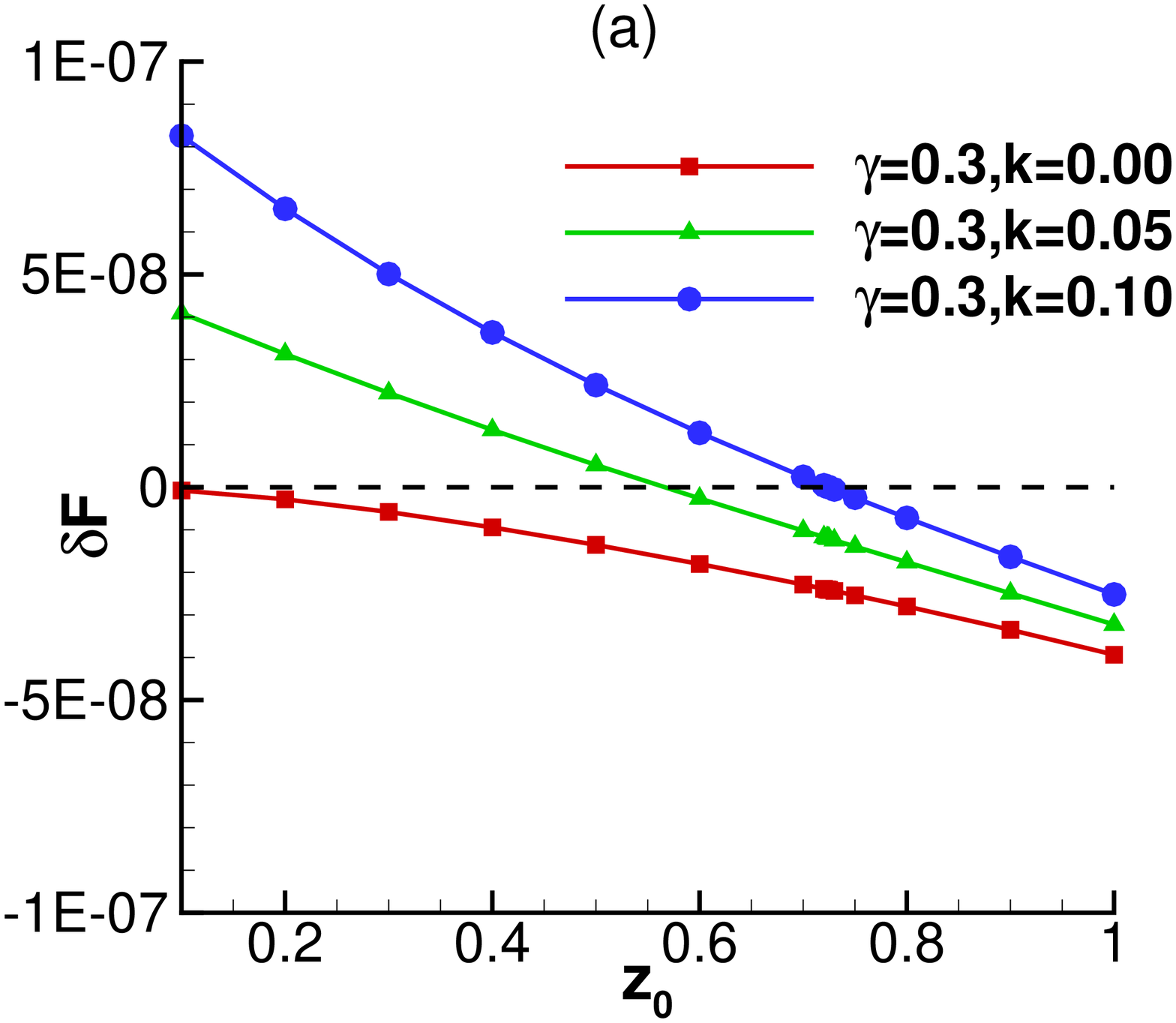}
\includegraphics[width=2.5in]{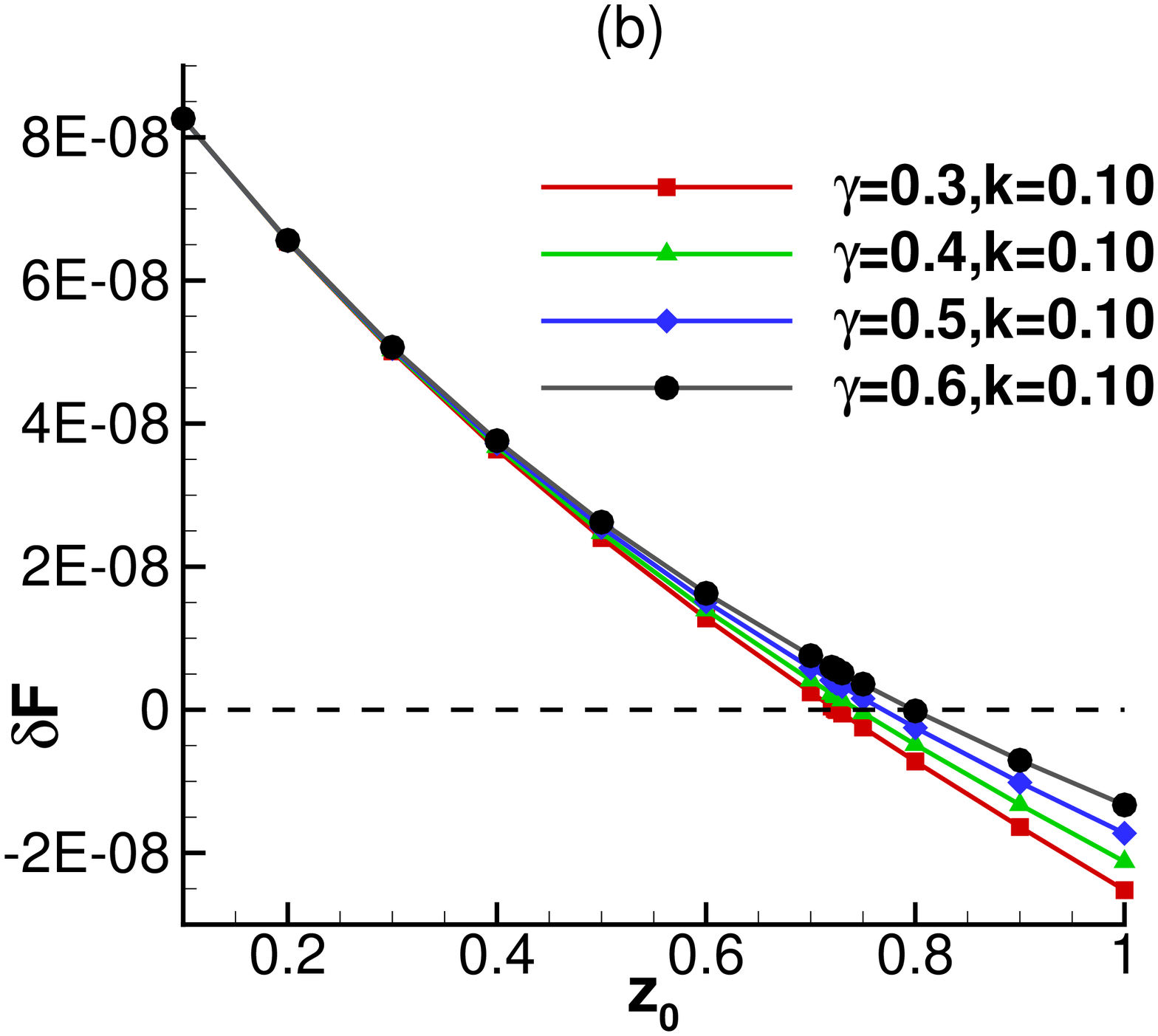}
\caption{The change $\delta F$ in the total free energy (Helfrich and edge energies) caused by a ribbon-like perturbation on the equilateral triangle protrusion as a function of the extent of stretching $z_0$ for different values of the mean curvature modulus $k$ and the line tension $\gamma$. (a) $\gamma=0.3$, $k=0.0,0.05$ and 0.10 from bottom to top. (b) $k=0.10$, $\gamma=0.3,0.4,0.5$ and 0.6 from bottom to top. The dashed horizontal line marks the line at $\delta F=0$ below which the perturbation is energetically favored.}
\label{Fig:BehPulTri}
\end{figure}

Turning to our second model of the protrusion, the equilateral triangle shown in Fig.~\ref{Fig:Protrusions}(b), we display in  Fig.~\ref{Fig:TermPulTri} the dependence of $B$, $A$ and $\delta L$ on the extent of stretching $z_0$, similar to Fig.~\ref{Fig:TermPul} for the Gaussian bump protrusion. The trends of the variations of these terms are similar to those in the Gaussian bump model. As in Fig.~\ref{Fig:TermPul}(b) we see from Fig.~\ref{Fig:TermPulTri}(b) that the integrated mean curvature squared curvature is nonzero for $z_0=0$, supporting our contention that the mean curvature leads to an energy barrier between disks and ribbons. Similar to Fig.~\ref{Fig:BehPul}, the change in the total energy $\delta F$ for the equilateral triangle protrusion is plotted in Fig.~\ref{Fig:BehPulTri} for a number of values of $k$ and $\gamma$.  For the triangle protrusion we find that $\delta F$  becomes negative when $z_{0c}=0.724\pm0.001$ for $k=0.1$ and $\gamma=0.3$ which is close to the value $0.569 \pm0.001$ found for the Gaussian bump.

\section{CONCLUSION}
\label{conclusions}

We have studied two kinds of instabilities of flat Sm--\textit{A}$^*$ monolayers with respect to the formation of twisted ribbons. The first instability is related to the phase transition from flat membranes to twisted ribbons which occurs when the concentration of depletant polymer (the edge energy modulus $\gamma$ in our theory) is lowered. We studied this transition using a model \cite{MeyerDiscussion} of the rippled disks which are structures that have been experimentally observed as precursors of the transition to ribbons.  Minimizing the energy of the model shape we found that the size of the ripples remains very small when the line tension is high (i.e., in the flat disk phase) while the size in the radial direction abruptly becomes large in a narrow range of $\gamma$ when $\gamma$ is lowered. This result is consistent with experimental observations of the growth of the ripples and the rapid increase in the radial size $A_r$ as compared to the imperceptible growth in the out of plane height $A_z$. Ripples with nonzero $A_z$ are regions of relatively high negative Gaussian curvature and we speculate that the rapid growth of $A_r$ with an accompanying nonzero value of $A_z$ is a signal of the instability to the formation of twisted ribbons.  The second instability we studied occurs when a membrane is stretched using optical traps. We studied this phenomenon by considering a ribbon-like perturbation added to a protrusion created at the edge of a  flat membrane by the optical trap. Our analysis of this instability was restricted to a free energy model which ignores the director field. Assuming that the mean curvature modulus $k$ is nonzero we found that the ribbon-like perturbation is energetically favorable once the protrusion reaches a nonzero critical  size $z_{0c}$. This result agrees qualitatively with experiment, however, our value of the critical stretching is an order of magnitude less than that observed experimentally, possibly due to our neglect of the director field.

For both instabilities we have found that the mean curvature energy acts as a barrier to the creation of twisted ribbons. In the case of the rippled disk we found that if $k \gtrsim 0.2$, the membrane remains flat and ripples are not energetically favorable.  In the case of stretching the critical value $z_{0c}$ grows with increasing $k$ until $k \sim 0.6$ and $z_{0c}\sim 1.3$. For larger values of $k$ the ribbon-like perturbation is no longer favored no matter how large we make $z_{0c}$.

\begin{acknowledgments}

We thank E. Barry, Z. Dogic, T. Gibaud, C. N. Kaplan, R. B. Meyer, P. Sharma and M. Zakhary for helpful discussions. We are grateful to R. B. Meyer for suggesting the model for the rippled disks studied here. This work was supported by the NSF through MRSEC Grant No. 0820492.

\end{acknowledgments}


\begin{thebibliography}{9}
\bibitem{footnote on fixed BC} We have also explored fixed boundary conditions for the director field. In that case we did not find the instability in $A_r$ found for free boundary conditions.
\end{thebibliography}
\end{document}